\documentclass[preprint,a4paper]{revtex4-1}
\usepackage[colorlinks,urlcolor=blue, citecolor=blue]{hyperref}
\usepackage[pdftex]{graphics,graphicx,colortbl,color,xcolor}
\usepackage[T1]{fontenc}
\usepackage{cancel}
\usepackage{ae,aecompl,amsthm,amsmath,amssymb,latexsym,fancyhdr,setspace,amscd}
\usepackage{wasysym}
\usepackage{verbatim} 

\usepackage[utf8x]{inputenc}

\begin{document}

\title{Running Vacuum Cosmological Models: Linear Scalar Perturbations}

\author{E. L. D. Perico}
\email{elduartep@usp.br}
\affiliation{Instituto de F\'{i}sica, Universidade de S\~ao Paulo, Rua do Mat\~ao 1371, CEP 05508-090, S\~ao Paulo, SP, Brazil}

\author{D. A. Tamayo}\email{tamayo@if.usp.br, tamayo.ramirez.d.a@gmail.com}
\affiliation{Departamento de Astronomia, Universidade de S\~ao Paulo, Rua do Mat\~ao 1226, 05508-900, S\~ao Paulo, Brazil}
\date{\today}

\begin{abstract}

In cosmology, phenomenologically motivated expressions for running vacuum are commonly parametrized as linear functions $\Lambda(H^2)$ or $\Lambda(R)$.
Such kind of models assume an equation of state for vacuum given by $\,\overline P_\Lambda=-\,\overline\rho_\Lambda$, relating their background pressure $\,\overline P_\Lambda$ and mean energy density $\,\overline\rho_\Lambda\equiv\Lambda/8\pi G$.
This equation of state requires that the dynamic for vacuum is due to the energy exchange with the material species.
Most of the approaches to background level consider only the energy exchange between vacuum and the transient dominant material component of the universe.
We extend such models assuming the running vacuum as the sum of independent contributions $\,\overline\rho_{\Lambda} =\sum_i\,\overline\rho_{\Lambda i}$, associated with (and interacting with) each of the $i$ material species.
We derive the linear scalar perturbations for two running scenarios, modeling its cosmic evolution and identifying their different imprints on the cosmic microwave background anisotropies and the matter power spectrum.
In the $\Lambda(H^2)$ scenario the running vacuum are coupled with all the material species in the universe, whereas the $\Lambda(R)$ description only leads to coupling between vacuum and the non-relativistic matter components; which produces different imprints of the two models on the matter power spectrum.
A comparison with the Planck 2015 data was made in order to constrain the free parameters of the models.
In the case of the $\Lambda(H^2)$ model, it was found that $\Omega_\Lambda=0.705\pm0.027$ and $H_0=69.6\pm2.9\, km\, Mpc^{-1}\, s^{-1}$, which diminish the tension with the low redshift expectations.

\end{abstract}

\maketitle
\onecolumngrid

\section{Introduction}

One of the most important discoveries of the 20th century is that the universe is expanding \cite{Hubble}, and more surprising is that it is accelerating \cite{RiessETAJ98,PerlmuterETNature98}.
At large scales gravity is the dominant force and, because its attractive nature in the context of General Relativity, the only way to have an accelerating universe is to assume a new cosmic component with the special feature to be gravitationally repulsive.
This component is known as dark energy (DE), and it must have some exotic characteristics like negative pressure, and permeate every part of the universe to have a global repulsive effect. 

The cosmological constant (CC) of the Einstein's field equations can account for such acceleration, and jointly with another unknown component referred as cold dark matter (CDM), turn out to be a remarkably good and widely accepted cosmological model known as $\Lambda$CDM \cite{BAO,WMAP,PlanckCosmology}.
However, this framework has some important theoretical problems.
In the general relativity context, a bare CC needs a fine-tuning of about 100 orders of magnitude, so that combined with the expected value of the vacuum energy density in quantum field theory, reproduce the effective dark energy density estimated from astronomical observations.
This theoretical conundrum is known as the CC problem \cite{WeinbergRMP89, Sahni:1999gb, Peebles:2002gy, Padmanabhan:2002ji, Sola:2013gha}.
Another hassle is that in spite of behaving quite differently with respect to the cosmic expansion the CDM and DE are found to contribute to the energy content of the universe today with amounts of the same order, this riddle is known as the cosmic coincidence problem.

Astronomical observations of different types support the existence of DE \cite{RiessETAJ98,PerlmuterETNature98, BAO, DWeinberg2013, WMAP, PlanckCosmology}, but do not provide a single clue about its origin from fundamental physics.
This allows the proposal for a wide range of types of DE candidates besides the pure positive CC like: quintessence, K-essence, chameleon field, $f(R)$ gravity and others; for a review see \cite{Yoo:2012ug,Joyce2015} and references therein.

Another dynamical DE formulation consider it as a decaying entity, which can be modeled as an effective interaction with the material components.
Since the lack of information on the nature of the DE it is difficult to describe these interactions from first principles; therefore, the interactions are often described phenomenologically.
The most worked approaches of interacting DE are the interacting dark sector models (DE and CDM), see \cite{Chimento2009, Koshelev2011, Abdalla} among many others.
In this vein, the model proposed in the present work extends this idea, taking into account the DE interact with all the other components, and assuming that DE can be separated into a sum of different contributions, each one associated only with one material component (say photons, baryons, CDM, etc.).

A recent motivation for the DE dynamics has emerged in the context of the renormalization group approach, where the simultaneous running of the CC and the gravitational coupling constant, due to quantum effects, has been considered \cite{BauerCQG05, ShapiroETJCAP05, GrandeETPLB07, GrandeETCQG10, Grande:2011xf}.
In those and in the present work, the \textit{running CC} is identified as the renormalized vacuum energy density.
Those renormalization group studies have shown that the corrections of the gravitational constant vary logarithmically with the scale of energy, therefore very slowly; while it is expected that the corrections to the CC are described by a power series.
That has led to increased interest in the study of CC running, leaving constant G \cite{Shapiro:2009dh, Sola:2013gha, SolaAIP14}, and this is the class of models in which we are interested.
As done by \cite{SolaAIP14} in the cosmological context, we identify the typical scale of energy of the process as the Hubble parameter, or the scalar curvature.
It is worth noting that before the renormalization group formulation of the CC running, decaying DE models were studied by several authors from the phenomenological point of view \cite{Lima1992, Overduin:1998zv}.

Most of the work done on this subject focuses mainly on the study of evolution, cosmological consequences and observational constraints of the running vacuum at the background level.
Effects of these models at perturbative level have been studied shallowly, but recently have begun to receive more attention \cite{Fabris07,Borges2008,Borges2008b,Borges2011,Basilakos2012,Toribio2012,borges2015,Gomez-Valent2015,Gomez-Valent2015b,GengLee2016,GengLee2016b}.
In such perturbative studies, the running vacuum is often modeled as decaying into the dominant material component of each cosmic era, i. e. without considering the contributions of the other components.
Thanks to the quantity and quality of current observational data, such an approximation may not be appropriate when modeling the evolution of linear perturbations.
However, expressions suggested by \cite{SolaAIP14} for the mean vacuum energy density of the form $\Lambda(H^2)$ and $\Lambda(R)$ lack a Lagrangian origin, the existence or the explicit form of the interaction to first order in perturbations is not given by itself.
The aim of the present work is to find a consistent formulation for the running vacuum perturbations and its material sources, as well to identify their observational imprints on the cosmic background radiation (CMB) and matter power spectrum.

The organization of the paper is as follows.
In section \ref{sec_renormalization_group}, we review two types of running vacuum models, one with $\Lambda(H^2)$ and another with $\Lambda(R)$.
In section \ref{sec_scalar_perturbations} we show the fluid conservation equations for coupled species, where the coupling terms still remain unspecified. 
In sections \ref{sec_L(H2)_pert_equations} and \ref{sec_L(R)_pert_equations}, we apply the linear scalar perturbation theory to the running vacuum models described in section \ref{sec_renormalization_group}. 
Using the Boltzmann equation, we find the coupling terms of the running vacuum with the material components for each model.
The behavior of the vacuum perturbations for sub-horizon modes is described, and the super-horizon initial conditions are founded.
In section \ref{sec_numerical_results} we show and discuss the result of integrating numerically the complete set of cosmological equations, for which was made use the free code CLASS \cite{class}.
Moreover, we use Planck 2015 data set \cite{PlanckData} and the statistical analysis package MontePython \cite{MontePython} to derive observational constraints.
Finally, in section \ref{sec_conclusions} we present our conclusions and some important remarks.

\section{Running Vacuum from Renormalization Group}\label{sec_renormalization_group}

In the cosmological context, several authors have motivated the time evolution of the DE density as a function of the Hubble parameter, $H$, using phenomenological arguments $\Lambda=\Lambda(H)$.
These models were confronted with observations -- supernovae, baryon acoustic oscillations (BAO), CMB, and large scale structure -- giving promising results \cite{Basilakos:2009wi,Grande:2011xf, Basilakos:2012ra, BasilakosMPA15, Sola:2015wwa, Gomez-Valent:2015pia, Fritzsch16,GengLee2016}.
In another approach, the CC problems have motivated the interest on the dynamical quantum effects on the vacuum energy density in quantum field theory, and their possible link and implications to DE concept in cosmology.
In this sense, the renormalization group formalism is used to study and parameterize the leading quantum effects of the vacuum energy in curved spacetimes (as an effective DE model), laying foundations for these models in more fundamental grounds \cite{Shapiro:2009dh, Sola:2011qr, Sola:2013gha, SolaAIP14}.

Usually, renormalization group approach in flat quantum field theory provide a useful theoretical tool to investigate how the gauge coupling constants and charges run with a scale $\mu$ associated to the typical energy of the process.
Similarly, the mean vacuum energy density $\overline\rho_\Lambda=\Lambda/8\pi G$ should depend on an energy scale of the gravitational processes on cosmological scales (G is the Newtonian gravitational constant).
The running of $\overline\rho_\Lambda$, due to quantum effects of the matter fields in the universe, can be associated with the change of the space-time curvature, and hence with the change of the typical energy of the classical gravitational external field in the Friedman-Lema\^itre-Robertson-Walker (FLRW) metric.
As this energy is pumped into the matter loops from the tails of the external gravitational field, it could be responsible for the physical running.
Using the renormalization group arguments, it is proposed that the term $\Lambda$ in the right hand side of the Einstein's equations can be expanded as a power series of $\mu$ \cite{SolaJPA08, Shapiro:2009dh, ShapiroSolaJHEP02, SolaAIP14}
\begin{equation}\label{RG eq}
\frac{d\,\Lambda}{d\,\mu} = m_0 +m_1\,\mu +m_2\,\mu^2+\cdots\,.
\end{equation}
This last expression is generic, and $\mu$ must be selected such that it properly traces the energy scale of the cosmological evolution.
In the following two subsections, we show two possible good selections for the energy scale $\mu$.

\subsection{$\Lambda(H^2)$ model}

A natural selection for the parameter $\mu$, on the running equation \eqref{RG eq} for $\Lambda$, consist in to fix it as the total energy density of the universe $\mu = \overline\rho_T \propto H^2$.
Therefore, we can rewrite eq. \eqref{RG eq} as: $\frac{d\,\Lambda}{d\,H^2} = c_2 +2\,c_4\,H^2 +\cdots +(n-2)\,c_n\,H^{n-2} +\cdots$ \cite{ShapiroSolaJHEP02, Sola:2013gha, PericoPRD13}, which can be integrated to obtain $\Lambda(H^2) =c_0 +c_2\,H^2 +c_4\,H^4 +\cdots +c_n\,H^n +\cdots$, where we have renamed conveniently the coefficients.

Because all the terms in the last expansion of $\Lambda$ must have the same dimension of $H^2$, the coefficients $c_n$ should have dimensions of $H^{n-2}$.
To choose the energy scale that fixes the dimensionality, instead of introducing a new energy scale, it is natural to choose the limit energy scale of the system where the theory still valid.
In order to be consistent with inflation, hence we choose to be the GUT scale $H_I^2$.
Consequently, the powers $H^n$ are suppressed by powers of $H_I^{n-2}$.
Accordingly, for the post-inflationary universe the terms of $\mathcal{O}(H^4)$ do not contribute effectively, and then, the last expression reduces to $\Lambda(H^2)\approx c_0+c_2\,H^2$.
Models $\Lambda \propto H^2$ have been studied in detail in \cite{Lima1992, ShapiroETPLB03, Basilakos:2012ra, Tamayo:2015qla}.
Whereas, the high order terms, which produce the inflationary stage, were studied by \cite{Tamayo:2015naa, LimaIJMPD15, BasliakosArxiv16} among others.

To adjust the constants we impose that the present vacuum energy density coincides with the measured today $\Lambda(H_0^2)\equiv \Lambda_0$, and for convenience, we define $c_2=3\alpha$.
These assumptions lead to a linear relation between $\Lambda$ and the total energy density ($H^2=8\pi G \,\overline\rho$):
\begin{equation}\label{H2_model}
\Lambda-\Lambda_0=3\alpha(H^2-H_0^2)\,,
\end{equation}
and can be rewritten as:
\begin{equation}\label{H2_split}
\,\overline\rho_\Lambda\equiv\rho_{\text{cr}}^0\,\frac{\Omega_\Lambda-\alpha}{1-\alpha}+\sum_i\rho_{\Lambda i}\,,
\end{equation}
where $\rho_{\text{cr}}^0$ is the present critical density, and we have defined
\begin{equation}\label{Lambda(H2)}
\,\overline\rho_{\Lambda i} \equiv \frac{\alpha}{1-\alpha}\,\overline\rho_i\,.
\end{equation}
We have divided the vacuum energy density into a sum of multiple contributions, one for each material component and related by the equation \eqref{Lambda(H2)}.
The index $i$ denotes each material component fluid, i.e., photons, neutrinos, massive neutrinos, CDM and baryons ($\gamma$, $\nu$, $h$, $c$, $b$); and the index $\Lambda i$ denotes the vacuum component associated with it.
The constant $\alpha$ is the model parameter for this case.

\subsection{$\Lambda(R)$ model}

Although $H^2$ realizes the total energy content of the universe, including spatial curvature, this is not a parameter that takes into account properly the 4-dimensional curvature scale of the universe.
In this sense, a most reasonable alternative for the energy scale is the scalar curvature $\mu=R$, the Ricci scalar.
Notice that a $\Lambda(R)$ model can be reduced to $\Lambda(H^2)$ model considering the flat FLRW metric in which $R=6(2H^2+\dot{H})$ and the special case $\dot{H}\approx 0$.
Substituting $\mu=R$ into equation \eqref{RG eq} we have $\Lambda(R) =c_0 +c_1\,R +c_2\,R^2 +\cdots +c_n\,R^n+\cdots$.
In this case, the coefficients $c_n$ of the expansion have dimension of $R_I^{1-n}$, where $R_I\approx 12 H_I^2$ is the typical curvature during the inflationary stage.
Once again, the high order terms do not contribute to the post-inflationary universe.
Setting $c_1=\beta$ (the model parameter for this case) we have that the vacuum energy density reduces to:
\begin{equation}\label{R_model}
\Lambda-\Lambda_0=\beta (R-R_0)\,.
\end{equation}
Models were $\Lambda \propto R$ have been studied in \cite{SolaAIP14}.
For the flat FLRW metric the scalar curvature is $R =8\pi G(\,\overline\rho_i -3\,\overline{\mathcal{P}}_i) + 4\Lambda$ then, we can rewrite
\begin{equation}\label{R_split}
\,\overline\rho_\Lambda\equiv \frac{\Lambda_0-\beta\,R_0}{8\pi G\,(1-4\beta)}+\sum_i\,\overline\rho_{\Lambda i}\,,
\end{equation}
where we have defined
\begin{equation}\label{Lambda(R)}
\,\overline\rho_{\Lambda i} \equiv \frac{\beta}{1 -4\beta}\,(\,\overline\rho_i -3\,\overline{\mathcal{P}}_i)\,.
\end{equation}
We have found two equations, \eqref{Lambda(H2)} and \eqref{Lambda(R)}, which relate the density of each material component $\overline\rho_i$ with its associated vacuum contribution $\overline\rho_{\Lambda i}$.
The parameters $\alpha$ and $\beta$ set the running rate of the vacuum.

Here are four important remarks to mention; first, we considered only the first two terms of the expansion ($H^2$ or $R$ plus a constant), because the contribution of high order terms is small after inflation.
Second, when the new parameters are null, we recover the usual CC values $\Omega_\Lambda = \,\overline{\rho}_{\Lambda_0}/\rho_{\text{cr}} = \Lambda_0/(3H^2_0)$ and $\Lambda=\Lambda_0$.
Third, as a working hypothesis, we decomposed the effective vacuum energy density as the sum of individual contributions associated with each of the material species. 
And fourth, as consequence, the running rate is the same for each material component.

\section{Scalar Linear Perturbations}\label{sec_scalar_perturbations}

In this section, we are going to summarize the cosmological scalar perturbations for pairs of fluids exchanging energy.
First, we show the usual conservation equations for a single fluid.
Then we will show the conservation equations for a pair of fluids, where one of them, the vacuum, has the equation of state $\,\overline P_\Lambda=-\,\overline\rho_\Lambda$, while the other has a non-negative pressure.
The last equations have a coupling term, which will be determined in the next section with the help of the Boltzmann equation applied to the material species.
The calculations and notation are based on the Ma and Bertschinger work \cite{Bertschinger}.

Let us consider a flat FLRW spacetime in the synchronous gauge, where the perturbed line element is:
\begin{equation}
ds^2 = a^2[-d\tau^2+(\delta_{ij}+h_{ij})dx^idx^j]\,,
\end{equation}
$\tau$ is the conformal time, $a=a(\tau)$ is the scale factor of the universe, and $h_{ij}=h_{ij}(\vec{x},\tau)$ is the metric perturbation, which can be written in the Fourier space as being
\begin{equation}
h_{ij}(\vec x,\tau)=\int d^3k\,e^{i\vec k\cdot \vec x}\left[\hat k_i\hat k_j\,h(\vec k,\tau)+\left(\hat k_i\hat k_j-\frac{\delta_{ij}}{3}  \right)\,6\,\eta(\vec k,\tau)  \right]\,,
\end{equation}
where $\vec k=k\hat k$, while $h$ and $\eta$ are the scalar functions that represent the scalar degrees of freedom of the perturbed metric.
In this gauge, the unperturbed Einstein field equations are:
\begin{equation}\label{Einstein0}
3\mathcal{H}^2 = 8\pi Ga^2\sum\,\overline\rho\,, \qquad
-2\frac{\ddot{a}}{a} +\mathcal{H}^2 = 8\pi Ga^2 \sum \overline{\mathcal{P}} \,,
\end{equation}
$\mathcal{H}\equiv\dot a/a$ is the comoving Hubble parameter, the overdot denotes derivative with respect to $\tau$, the overline denotes background averaged quantities, and the sum cover all energetic species (material components $i$ and vacuum counterparts $\Lambda i$).
To first order in this gauge, the evolution equations for scalar perturbations are:
\begin{subequations}\label{Einstein}
\begin{align}
\label{Einstein1}
k^2\eta-\frac{1}{2}\mathcal{H}\dot h =& -4\pi Ga^2\sum\,\overline\rho\,\delta\,,\\
\label{Einstein2}
-k^2\dot\eta =& -4\pi Ga^2\sum\left(\,\overline\rho+\overline{\mathcal{P}}\right)\theta\,,\\
\label{Einstein3}
\ddot h+2\mathcal{H}\dot h-2k^2\eta =& -8\pi Ga^2\sum \delta\mathcal{P}\,,\\
\label{Einstein4}
\left(\ddot h+6\ddot\eta\right)+2\mathcal{H}\left( \dot h+6\dot \eta \right)-2k^2\eta =& -24\pi Ga^2\sum\left( \,\overline\rho+\overline{\mathcal{P}} \right)\sigma\,,
\end{align}
\end{subequations}
where $\delta\equiv\delta\rho/\overline\rho$ is the density contrast, $k_ik_j\Sigma^i_j\equiv-k^2\left(\,\overline\rho+\,\overline{\mathcal{P}}\right)\sigma$ is the anisotropic stress, $(\overline\rho+\overline{\mathcal{P}})\theta\equiv ik^ i\delta T^ 0_j$, and $T^\mu_\nu\equiv\overline{T^\mu_\nu}+\delta T^\mu_\nu$ is the energy-momentum tensor, which, in the fluid form can be written as
\begin{equation}\label{stress-energy-fluid}
T^0_0=-\sum\left(\overline\rho+\delta\rho \right)\,, \qquad\qquad
T^0_j=\sum\left(\overline\rho+\overline{\mathcal{P}}\right)v_j\,, \qquad\qquad
T^k_j=\sum(\overline{\mathcal{P}}+\delta\mathcal{P})\delta^k_j+\Sigma^k_j\,,
\end{equation}
where $v_j$ is the peculiar 3-velocity of the fluid, with respect to the Hubble flow, and $\Sigma^k_j$ is the traceless anisotropic stress part of the energy-momentum tensor.
In addition, we will assume an equation of state in the form $\overline{\mathcal{P}}=\omega\overline\rho$ with $\omega=$const. for each of the fluid species. 

\subsection{Conservation equations for a single fluid}

For a single perfect fluid, the temporal part of the energy-momentum tensor conservation equation ($T^{\mu0}_{;\mu}$=0) becomes:
\begin{subequations}\label{dot_rho}
\begin{align}
\dot{\,\overline\rho}+3\mathcal{H}\,\overline\rho\left(1+\omega \right)=\,0\,,\\
\,\overline\rho\left(1+\omega  \right)\theta+3\mathcal{H}\left(\delta\mathcal{P}+\,\overline\rho\,\delta \right) +\,\overline\rho\,\dot\delta+\dot{\,\overline\rho}\,\delta +\,\overline\rho\left(1+\omega\right)\frac{\dot h}{2} =\, 0 \,,
\end{align}
\end{subequations}
the first for the background evolution of the energy density, and the second for the density contrast evolution.
In the same way, the spatial component ($T^{\mu i}_{;\mu}=0$) gives to us an evolution equation for $\theta$:
\begin{equation}\label{dot_theta}
\dot{\,\overline\rho}(1+\omega)\theta+\,\overline\rho\left(1+\omega \right)\dot \theta+4\mathcal{H}\,\overline\rho\left(1+\omega \right)\theta-k^2\delta\mathcal{P}+k^2\,\overline\rho\left(1+\omega \right)\sigma =0 \,.
\end{equation}

\subsection{Conservation equation for a vacuum-material coupled pair}

As we said before, we model the total energy density of the running vacuum (Eq. \eqref{H2_model} or \eqref{R_model}) splitting it into a collection of partial vacuum components each one associated with only one material counterpart (Eq. \eqref{H2_split} or \eqref{R_split}).
The coupled pairs are composed by each of the vacuum components $\Lambda i$ and it associated material source $i$.
We assume that the background total energy density of vacuum satisfy the equation of state $\,\overline\rho_\Lambda+\,\overline{\mathcal{P}}_\Lambda=0$, as also each partial component $\,\overline\rho_{\Lambda i}+\overline{\mathcal{P}}_{\Lambda i}=0$.
Therefore, the conservation equation, and eqs. \eqref{dot_rho} and \eqref{dot_theta}, for the coupled pair $(\Lambda i,i)$ are
\begin{subequations}\label{dot_split}
\begin{align}\label{dot_splita}
\dot{\,\overline\rho}_{\Lambda i}=&-Q_{0i}\,,\\\label{dot_splitb}
\dot{\,\overline\rho}_i+3\mathcal{H}\left(\,\overline\rho_i+\,\overline{\mathcal{P}}_i\right)=&\,\quad Q_{0i}\,,\\\label{dot_splitc}
3\mathcal{H}\left(\delta\mathcal{P}_{\Lambda i}+\delta\rho_{\Lambda i} \right) +\,\overline\rho_{\Lambda i}\,\dot\delta_{\Lambda i}+\dot{\,\overline\rho}_{\Lambda i}\,\delta_{\Lambda i}=&-Q_{1i}\,,\\\label{dot_splitd}
\left(\,\overline\rho_i+\,\overline{\mathcal{P}}_i\right)\theta_i+3\mathcal{H}\left(\delta\mathcal{P}_i+\delta\rho_i \right) +\,\overline\rho_i\,\dot\delta_i+\dot{\,\overline\rho}_i\,\delta_i +\left(\,\overline\rho_i+\,\overline{\mathcal{P}}_i\right)\frac{\dot h}{2}=&\,\quad Q_{1i}\,,\\\label{dot_splite}
-k^2\delta\mathcal{P}_{\Lambda i}=&-Q_{2i}\,,\\\label{dot_splitf}
\left(\dot{\,\overline\rho}_i+\dot{\,\overline{\mathcal{P}}}_i\right)\theta_i+\left(\,\overline\rho_i+\,\overline{\mathcal{P}}_i\right)\dot \theta_i+4\mathcal{H}\left(\,\overline\rho_i+\,\overline{\mathcal{P}}_i\right)\theta_i -k^2\delta\mathcal{P}_i+k^2\left(\,\overline\rho_i+\,\overline{\mathcal{P}}_i\right)\sigma_i=&\,\quad Q_{2i}\,.
\end{align}
\end{subequations} 
These are the evolution equations for background densities, pressures, and perturbations for each pair composed of a material component and its corresponding vacuum counterpart.
Each equation was divided into two parts, one for the material fluid and the other for the associated vacuum,.
They are linked by  coupling terms $Q$'s to be determined.
To solve the evolution equations of the perturbations \eqref{dot_split}, we should know explicitly the form of the couple terms $Q$'s.

Note that the equation \eqref{dot_splite} corresponds to a ligature for $\delta\mathcal{P}_{\Lambda i}$, which is the result of the vacuum energy-momentum tensor parameterization in the usual fluid version \eqref{stress-energy-fluid}.
A more general case would be not to use any prescription for $T^{0i}$ component, in which case it would emerge a dynamic equation to this perturbation; but in return, we would have to have an external expression for $\delta\mathcal{P}_{\Lambda i}$.

In the particular case of $\,\overline{\mathcal{P}}_i/\,\overline\rho_i=\omega_i=$const. the equations (\ref{dot_splitc}-\ref{dot_splitf}) can be written as
\begin{subequations}\label{dot_split_}
\begin{align}\label{dot_splita_}
3\mathcal{H}\left(\frac{Q_{2h}}{k^2}+\delta\rho_{\Lambda i} \right) +\,\overline\rho_{\Lambda i}\,\dot\delta_{\Lambda i}=&-Q_{1i}+Q_{0i}\,\delta_{\Lambda i}\,,\\\label{dot_splitb_}
\rho_i\left(1+\omega_i\right)\theta_i + 3\mathcal{H}\left(\delta\mathcal{P}_i- \omega_i\,\delta\rho_i\right) +\,\overline\rho_i\,\dot\delta_i +\rho_i\left(1+\omega_i\right)\frac{\dot h}{2}=&\,\quad Q_{1i}-Q_{0i}\,\delta_i\,,\\\label{dot_splitc_}
\rho_i\left(1+\omega_i\right)\dot \theta_i+\mathcal{H}\rho_i\left(1-3\omega_i\right)\theta_i -k^2\delta\mathcal{P}_i+k^2\rho_i\left(1+\omega_i\right)\sigma_i=&\,\quad Q_{2i}-Q_{0i}\left(1+\omega_i\right)\theta_i\,.
\end{align}
\end{subequations}

We want to highlight the fact that we will only use these equations which correspond to vacuum components $\Lambda i$.
Both the coupling terms $Q$'s, as the equations of evolution of material components $i$ will be obtained from the Boltzmann formalism.
This will be done separately for each running model in the next two sections.

\section{$\Lambda(H^2)$ Perturbations}\label{sec_L(H2)_pert_equations}

We assume that the energy-momentum tensor of each coupled pair $\{i,\Lambda i\}$ (a given material species and its vacuum partner) is conserved independently of the other pairs $\{j,\Lambda j\}$, further, it can be separated in two individual equations by using a common coupling term, for details see section \ref{sec_scalar_perturbations}.
The goal of this section is to model the coupling terms between vacuum perturbations and their material counterparts.
As a bonus, we will found that the two coupling terms, in the right-hand side of equations \eqref{dot_splitb_} and \eqref{dot_splitc_}, are canceled between them, leaving the evolution equations for material perturbations, of the running models, exactly like $\Lambda$CDM equations.
Thus, material perturbations only are affected by the running indirectly, through the metric source $h$ and the background quantities (both modified directly by the running vacuum).

The zero order coupling term $Q_{0i}$ can be obtained as follows.
Using the expression \eqref{Lambda(H2)} for $\bar\rho_{\Lambda i}=\bar\rho_{\Lambda i}(\bar\rho_i)$, which defines our split $\bar\rho_\Lambda=const.+\sum_i\bar\rho_{\Lambda i}$ such that $\Lambda-\Lambda_0=3\alpha(H^2-H_0^2)$, into the background conservation equations (\ref{dot_splita} and \ref{dot_splitb}), we can express the previous two as
\begin{equation}\label{conservacion_0}
\dot{\,\overline\rho}_i+3\mathcal{H}\left(\,\overline\rho_i+\,\overline{\mathcal{P}}_i \right)=3\alpha\mathcal{H}\left(\,\overline\rho_i+\,\overline{\mathcal{P}}_i \right)\equiv Q_{0i}\,,
\end{equation}
which have an analytical solution when $\,\overline{\mathcal{P}}_i/\,\overline\rho_i = \omega_i =$const.:
\begin{equation}\label{densities_H2}
{\,\overline\rho}_i={\,\overline\rho}_i^0\,a^{-3(1-\alpha)(1+\omega_i)}\,.
\end{equation}

At this point, the fluid approximations and the $\Lambda(H^2)$ background expression, which define the model studied here, do not provide direct information about the coupling terms $Q_{1i}$ and $Q_{2i}$ for perturbations.
We will use the Boltzmann equation applied to material species in order to find this coupling terms, which will be used in the fluid equations for the vacuum species.

First, we look for the collision term of the Boltzmann equation that reproduces the zero order equation \eqref{conservacion_0}; and then, we will use the natural extension of that collision term for calculate the coupling terms $Q_{1i}$ and $Q_{2i}$ for linear perturbations.

In addition, the Boltzmann formalism becomes unavoidable when dealing with relativistic species such as photons ($\gamma$), massless neutrinos ($\nu$) and massive neutrinos $(h$).
This is because the fluid approximation does not provide an equation for the anisotropic stress $\sigma$ evolution; and does not take into account higher order multipoles, which are coupled to low order multipoles line $\delta$ and $\theta$.

Therefore, the goal of this section will be to use the Boltzmann formalism --applied in this work only for the material species-- for extend the zero order collision term, which reproduces the background dynamic of the model, to the first order level of perturbations.
We will apply these collision terms to the fluid equations that describe the vacuum species.

\subsection{Coupling terms for massive neutrinos}

The energy-momentum tensor can be expressed in terms of the phase-space distribution function $f(x^i,P_j,\tau)$ as follow:
\begin{equation}
T_{\mu\nu}=\int dP_1\,dP_2\,dP_3\,(-g)^{-1/2}\frac{P_\mu\,P_\nu}{P_0}f(x^i,P_j,\tau)
\end{equation}
where $g$ is the trace of the metric, and $P^\mu$ is the 4-momentum of the particles of a given species.
The conjugate momentum $P^i$ can be expressed in the syncrhonous gauge as a function of the proper momentum $p^ i=p_i$ as $P_i=a(\delta_{ij}+h_{ij}/2)p^j$, and in turn we can define a comoving 3-momentum $q_j\equiv ap_j=qn_j$, where $n^in_j=1$. 

The phase space distribution function evolves according to the Boltzmann equation:
\begin{equation}\label{Boltzmann}
\frac{D\,f}{d\tau}=\frac{\partial f}{\partial\tau}=\frac{dx^i}{d\tau}\frac{\partial\,f}{\partial x^i}+\frac{dq}{d\tau}\frac{\partial\,f}{\partial q}+\frac{dn_i}{d\tau}\frac{\partial\,f}{\partial n_i}=\left(\frac{\partial\,f}{\partial\tau}  \right)_C\,,
\end{equation}
where the right hand side of the equation represents the collision term, such as the number of particles $N$
\begin{equation}
f\left( x^i,P_j,\tau \right)dx^1\,dx^2\,dx^3\,dP_1\,dP_2\,dP_3=dN\,,
\end{equation}
is conserved is the collision term vanishes.

Following the Ma and Bertschinger procedure in \cite{Bertschinger}, we will express the phase-space distribution function $f(x^i,P_j,\tau)$ as a the sum of a zero-order expression, $f_0(q,\tau)$, plus a linear perturbation parameterized as
$
f(x^i,P_j,\tau)=f_0(q,\tau)\left[1+\Psi(x^i,q,n_j,\tau)  \right]\,
$.
Replacing this expression into the Boltzmann equation \eqref{Boltzmann}, the last can be split into its zero and first order components:
\begin{subequations}
\begin{align}
\label{dot_f0}
\frac{\partial f_0}{\partial \,\tau}=&\left(\frac{\partial f}{\partial \tau}\right)_{C 0}\,,\\
\label{dot_Psi}
\frac{\partial \ln f_0}{\partial\,\tau}\Psi+\frac{\partial \Psi}{\partial \tau}+i\,\frac{q}{\epsilon}(\vec k\cdot \hat n)\,\Psi+\frac{d\,\ln\,f_0}{d\,\ln\,q}\left[\dot \eta-\frac{\dot h+6\dot \eta}{2}\left( \vec k\cdot \hat n \right)^2  \right]=&\frac{1}{f_0}\left(\frac{\partial f}{\partial \tau}\right)_{C 1}\,,
\end{align}
\end{subequations}
where $\epsilon=\sqrt{q^2+a^ 2 m^ 2}=\sqrt{P^ 2+a^ 2m^ 2}=-P_0$, and $m$ is the mass of the particles.
Expanding the field $\Psi$ in Legendre series
\begin{equation}
\Psi(\vec k,\hat n,q,\tau)\equiv\sum\limits_{l=0}^\infty(-i)^l(2l+1)\Psi_l(\vec k,q,\tau)P_l(\hat k\cdot\hat n)\,.
\end{equation}
and substituting it into the Boltzmann equation \eqref{dot_Psi} are obtained the differential equations for the evolution of the fields $\Psi_l$.
Integrating out the $q$ dependence it can re-obtain the components of the energy-momentum tensor:
\begin{eqnarray}
\,\overline\rho_h=\,4\pi a^{-4}\int q^2\,dq\,\epsilon\,f_0(q,\tau)\,, &\quad
\,\overline{\mathcal{P}}_h=\,\frac{4\pi}{3}a^{-4}\int q^2\,dq\,\frac{q^2}{\epsilon}\,f_0(q,\tau)\,,\nonumber\\
\delta\rho_h=\,4\pi a^{-4}\int q^2\,dq\,\epsilon\,f_0(q,\tau)\,\Psi_0\,,&\quad
\delta\mathcal{P}_h=\,\frac{4\pi}{3} a^{-4}\int q^2\,dq\,\frac{q^2}{\epsilon}\,f_0(q,\tau)\,\Psi_0\,,\\
\left(\,\overline\rho_h+\,\overline{\mathcal{P}}_h  \right)\theta_h=\,4\pi k a^{-4}\int q^2\,dq\,q\,f_0(q,\tau)\,\Psi_1\,,&\quad
\left(\,\overline\rho_h+\,\overline{\mathcal{P}}_h  \right)\sigma_h=\,\frac{8\pi}{3} a^{-4}\int q^2\,dq\,\frac{q^2}{\epsilon}\,f_0(q,\tau)\,\Psi_2\,.\nonumber
\end{eqnarray}
In order to reproduce the background conservation equation \eqref{conservacion_0}, the zero order distribution function must satisfy the follow equation:
\begin{equation}\label{dot_f0_h}
\dot f_0=3\alpha\mathcal{H}\left(1+\frac{q^2}{3\epsilon^2}\right)f_0\,,
\end{equation}
whose time evolution is due to the interaction with his vacuum counterpart.
This distribution function $f_0(q,\tau)$ can be related to the \textit{free} neutrino distribution function $\tilde f_0(q)$ as:
\begin{equation}\label{f_0(q,tau)}
f_0(q,\tau)\equiv a^{4\alpha}\left(\frac{\epsilon_0}{\epsilon}\right)^\alpha \tilde f_0(q)\,,\qquad\qquad\text{where}\qquad\qquad\tilde f_0=\tilde f_0(\epsilon)=\frac{g_s}{e^ {q/k_BT_0}\pm1}\,.
\end{equation}
$g_s$ is the number of spin degrees of freedom, $T_0$ is the temperature of the particles today, $k_B$ is the Boltzmann constant, and $\epsilon_0=\epsilon(a=1)$.
Based on the code CLASS work \cite{class_neutrinos}, which we use for numerical integration, the \textit{free} neutrino distribution function, $\tilde f_0(q)$, is modeled here as warm dark matter, which is time independent.
Other massive neutrino formulations in the literature include the Degenerate Fermion Gas approximation, see \cite{BernardiniJCAP11, PericoJCAP11}, while a time-dependent distribution function was discussed by \cite{neutrino2013}. 
The collision term of the Boltzmann equation that reproduce the expression \eqref{dot_f0_h} is
\begin{equation}\label{neutrino_collission_term}
\left(\frac{\partial f}{\partial \tau}\right)_C=3\alpha\mathcal{H}\left(1+\frac{q^2}{3\epsilon^2}\right)\,f\,,
\end{equation}
and the coupling terms derived therefrom are
\begin{subequations}\label{Q_massive_neutrinos}
\begin{align}
Q_{0 h}=&\,4\pi a^{-4}\int q^2\,dq\,\epsilon\,\dot f_0(q,\tau)\quad\;=3\alpha\mathcal{H}\left(\,\overline\rho_h+\,\overline{\mathcal{P}}_h\right)\,,\\
Q_{1 h}=&\,4\pi a^{-4}\int q^2\,dq\,\epsilon\,\dot f_0(q,\tau)\,\Psi_0=3\alpha\mathcal{H}\left(\delta\rho_h+\delta\mathcal{P}_h\right)\,,\\
Q_{2 h}=&\,4\pi k a^{-4}\int q^3\,dq\,\dot f_0(q,\tau)\,\Psi_1\nonumber\\
=&\,3\alpha\mathcal{H}\left(\,\overline\rho_h+\,\overline{\mathcal{P}}_h\right)\theta_h+3\alpha\mathcal{H}\,4\pi k a^{-4}\int q^3\,dq\,\frac{q^2}{3\epsilon^2}\,f_0(q,\tau)\,\Psi_1\,.
\end{align}
\end{subequations}
Substituting this coupling terms on the fluid evolution equation \eqref{dot_splita_}, for the perturbations of the vacuum component associated to massive neutrinos, we obtain
\begin{equation}\label{dot_delta_lambda_massive_neutrinos}
\,\overline\rho_{\Lambda h}\left(\dot\delta_{\Lambda h} + 3\mathcal{H} \delta_{\Lambda h}\right) +3\mathcal{H}\frac{Q_{2 h}}{k^2}=3\alpha\mathcal{H}\left[\left(\,\overline\rho_h+\,\overline{\mathcal{P}}_h\right)\delta_{\Lambda h}-\left(\delta\rho_h+\delta\mathcal{P}_h\right) \right]\,.
\end{equation}
On the other hand, notice that substituting the collision term \eqref{neutrino_collission_term} (proportional to the distribution function) into the first order Boltzmann equation \eqref{dot_Psi}, for the massive neutrinos, we obtain a collisionless form
\begin{equation}
\frac{\partial \Psi}{\partial \tau}+i\,\frac{q}{\epsilon}(\vec k\cdot \hat n)\,\Psi+\frac{d\,\ln\,f_0}{d\,\ln\,q}\left[\dot \eta-\frac{\dot h+6\dot \eta}{2}\left( \vec k\cdot \hat n \right)^2  \right]=0\,.
\end{equation}
Therefore, the evolution equations for material perturbations remain unchanged, not only for massive neutrinos, since our only demand was really that $\dot{\tilde f}_0=0$.
In particular, substituting the collision term \eqref{neutrino_collission_term} in the Boltzmann equation we obtain that the evolution equations for the multipoles $\Psi_l$ are the same as those of the reference \cite{Bertschinger}, the only difference is that for the our model $f_0(q,\tau)$ is related to $\tilde f_0(q)$ (denoted as $f_0(q)$ in \cite{Bertschinger}) by the eq. \eqref{f_0(q,tau)}, and then:
\begin{equation}
\frac{\partial\ln f_0}{\partial\ln q}=\frac{\partial\ln\tilde f_0}{\partial\ln q}+\alpha\,\left(\frac{q^2}{\epsilon_0^2}-\frac{q^2}{\epsilon^2}\right)\,.
\end{equation}

\subsection{Coupling terms for energetic components with $\omega=$constant}

In the case of $\omega_i =\,\overline{\mathcal{P}}_i/\,\overline\rho_i =\delta\mathcal{P}_i/\delta\rho_i =const.$ the analysis made in the last section reduces significantly.
In order to recover for the evolution of the mean matter density \eqref{conservacion_0}, for the material species with $\omega_i\equiv\overline{\mathcal{P}}_i/\overline{\rho}_i$, we have that the collision term must be equal to
\begin{equation}\label{collision_term}
\left(\frac{\partial\,f}{\partial\tau}  \right)_C=3\,\alpha\,(1+\omega_i)\,\mathcal{H}\,f\,,
\end{equation}
which leads to the coupling terms
\begin{subequations}\label{collision Q terms H2}
\begin{align}\label{collision Q terms H2a}
Q_{0i}=&\;3\alpha(1+\omega_i)\mathcal{H}\,\overline\rho_i\,,\\\label{collision Q terms H2b}
Q_{1i}=&\;3\alpha(1+\omega_i)\mathcal{H}\,\overline\rho_i\delta_i\,,=Q_{0i}\,\delta_i\,,\\\label{collision Q terms H2c}
Q_{2i}=&\;3\alpha(1+\omega_i)^2\mathcal{H}\,\overline\rho_i\theta_i\,=Q_{0i}\,(1+\omega_i)\theta_i\,.
\end{align}
\end{subequations}
Remarkably, the last collision terms makes the matter Boltzmann equations to first order become the same as in \cite{Bertschinger}, because the two terms in the right hand side cancel in Eqs. (\ref{dot_splitb_}, \ref{dot_splitc_}); while Eq. \eqref{dot_splita_} for the vacuum perturbation $\Lambda i$ becomes:
\begin{equation}\label{dot_delta_lambda}
\dot\delta_{\Lambda i}+3\mathcal{H}\delta_{\Lambda i}+\frac{9(1+\omega_i)^2\mathcal{H}^2}{k^2}(1-\alpha)\theta_i=3(1+\omega_i)(1-\alpha)\mathcal{H}\left( \delta_{\Lambda i}-\delta_i\right)\,,
\end{equation}
where $\delta_{\Lambda i} = \delta\rho_{\Lambda i}/\,\overline{\rho}_{\Lambda i}$ and $\delta_i = \delta\rho_i/\,\overline{\rho}_i$ are the density contrasts.
In this limit, $\omega_i=\overline{\mathcal{P}}_i/\overline\rho_i=const.$, expressions \eqref{Q_massive_neutrinos} and \eqref{dot_delta_lambda_massive_neutrinos} reduce to \eqref{collision Q terms H2} and \eqref{dot_delta_lambda}, respectively.

\subsection{Super-horizon initial conditions}\label{super_H2}

To follow expressions summarize initial conditions, to the equations \eqref{dot_delta_lambda_massive_neutrinos} and \eqref{dot_delta_lambda} for vacuum perturbations and usual material equations \cite{Bertschinger}, during the radiation-dominated era and super-horizon scales, $k\ll\mathcal{H}$:
\begin{subequations}
\begin{align}
\delta_{\Lambda c}=&\,\frac{3}{2}\frac{1-\alpha}{2-\alpha}\,C_1\,\left(k\,\tau \right)^2\,,\\
\delta_{\Lambda\gamma}=&\,\frac{1}{16}\,\frac{(1-\alpha)(1-3\alpha)}{1-2\alpha}\,C_1\,\left(k\,\tau \right)^2\,,\\
\delta_{\Lambda b}=&\frac{2-5\alpha+3\alpha^2}{(2-\alpha)(1-2\alpha)}\,C_1\,\left( k\,\tau \right)^2\,,\\
\delta_{\Lambda\nu}=&\frac{16}{9}\,\frac{34+8R-\alpha(45+12R_\nu)}{15+4R_\nu}\,\frac{1-\alpha}{1-2\alpha}\,C_1\,\left(k\,\tau \right)^2\,,\\
\delta_{\Lambda h}=&\delta_{\Lambda\nu}\,,\\
\sigma_\nu=&\frac{R_\nu(6-8\alpha)-15\alpha}{3R_\nu(15+4R_\nu)}\,\frac{1-2\alpha}{3+2\alpha}\,2\,C_1\,(k\,\tau)^2\,,
\end{align}
\end{subequations}
other initial conditions are the same as those in \cite{Bertschinger}.
It was used $\rho_T\approx \rho_\gamma+\rho_\nu+\rho_{\Lambda\gamma}+\rho_{\Lambda\nu}$, and $\rho_\gamma = \rho_\gamma^0 a^{-4(1-\alpha)}$ and $\rho_\nu = \rho_\nu^0a^{-4(1-\alpha)}$, which leads to
\begin{equation}\label{a H2}
a^{1-2\alpha}=\tau\,(1-2\alpha)\sqrt{\cfrac{8\pi G(\,\overline\rho^0_\nu+\,\overline\rho^0_\gamma)}{3(1-\alpha)}}\,,\qquad \text{with} \qquad\mathcal{H}=\frac{1}{\tau\left(1-2\alpha\right)}\,.
\end{equation}
$R_\nu$ was defined as $R_\nu \equiv\,\overline\rho_\nu/(\,\overline\rho_\nu+\,\overline\rho_\gamma)$, which is a constant ratio during the radiation dominated era regardless the value of $\alpha$.
All initial conditions reduce to the $\Lambda$CDM case when $\alpha=0$.

\subsection{Sub-horizon evolution for vacuum perturbations}\label{sub_H2}

For sub-horizon modes, the evolution equation \eqref{dot_delta_lambda} for the density contrast of the vacuum components associated to the non-relativistic material species becomes:
\begin{equation}
a\,\delta_{\Lambda}'=-3\alpha\delta_{\Lambda} -3(1-\alpha)\delta\,,
\end{equation}
where the line means derivative with respect to the scale factor $a$.
Most of the cosmological constraints for this kind of models show that $\alpha$ must be in the range of $10^{-5}-10^{-3}$ \cite{SolaAIP14, Gomez-Valent:2015pia, Sola:2015wwa}, which let us estimate $\delta_\Lambda$ as being around $\delta_\Lambda\sim-3(1-\alpha)\delta$.
The bigger amplitude of the density contrast for the vacuum components, in comparison with they material counterparts, does not represent an observational problem for the model, since such perturbations come into Einstein's equations of the form of $\delta\rho_\Lambda=\,\overline\rho_\Lambda\,\delta_\Lambda\approx\alpha\,\overline\rho\,\delta_\Lambda$, which is suppressed by a factor $\alpha$.

For the vacuum components associated with the relativistic matter, the equation \eqref{dot_delta_lambda} becomes:
\begin{equation}
a\,\delta_\Lambda'\approx \delta_\Lambda (1-4\alpha)-4\delta(1-\alpha)\,.
\end{equation}
In addition, the density contrast of the relativistic material components decrease considerably inside the horizon, therefore, the last term of the above equation can be removed.
In this case, the vacuum perturbation evolves as $\delta_\Lambda\propto a^ {1-4\alpha}$, similar to the linear growth of the non-relativistic perturbations.

During the tight coupled baryon-photon stage, the equation for perturbation of the vacuum fraction associated to baryons becomes:
\begin{equation}
a\,\delta_\Lambda'=-3\alpha\delta_\Lambda-3\delta(1-\alpha)\,.
\end{equation}
During this stage, the amplitude of the tightly coupled baryon and photon density contrasts decrease significantly.
Furthermore, because the small value expected for $\alpha$, the right-hand side of the last equation can be neglected, which gives $\delta_\Lambda\approx$const.

Finally, the last equation is also valid for the vacuum component associated to massive neutrinos after the non-relativistic transition, where $\delta_{\Lambda}\ll\delta$ as was inferred above, and since $\alpha\ll1$ we obtain $\delta_\Lambda\approx$const.

\section{$\Lambda(R)$ Perturbations}\label{sec_L(R)_pert_equations}

In this section, we are going to apply the same procedure done in the previous section for the running vacuum  model of the type $\Lambda = \Lambda_0 + \beta(R-R_0)$.
The expression \eqref{Lambda(R)}, which defines our split of the $\Lambda(R)$ model, let us rewrite the background conservation equations (\ref{dot_splita}, \ref{dot_splitb}) as follow
\begin{equation}\label{conservacion_0_R}
(1-3\beta)\dot{\,\overline\rho}_i-3\beta\dot{\,\overline{\mathcal{P}}}_i+3(1-4\beta)\mathcal{H}(\,\overline\rho_i+\,\overline{\mathcal{P}}_i)=0\,.
\end{equation}

\subsection{Coupling terms for massive neutrinos}

In order to reproduce the evolution equation for the massive neutrinos mean density \eqref{conservacion_0_R}, his zero order distribution function must satisfy the following equation:
\begin{equation}\label{dot_f0_R}
\dot f_0=3\beta\mathcal{H}\,\frac{\epsilon^2}{\varepsilon^2}\left(1-\frac{q^2}{3\epsilon^2}\right)\left(1-\frac{q^2}{\epsilon^2}\right)f_0\,.
\end{equation}
This distribution function $f_0(q,\tau)$ can be related to the `free' neutrino distribution function $\tilde f_0(q)$ as:
\begin{equation}\label{f_0(q,tau)_R}
f_0(q,\tau)\equiv \frac{\epsilon}{\epsilon_0}\,\left(\frac{\varepsilon_0}{\varepsilon}\right)^{(1-6\beta)/(1-3\beta)} \tilde f_0(q)\,,\qquad\qquad\text{such that}\qquad\qquad \dot{ \tilde f}_0(q)=0\,,
\end{equation}
where $\epsilon_0=\epsilon(a=1)$, $\varepsilon_0=\varepsilon(a=1)$ and
\begin{equation}
\varepsilon(q,a)\equiv\sqrt{q^2(1-4\beta)+m^2a^2(1-3\beta)}\,.
\end{equation}
The collision term for the Boltzmann equation that reproduces the expression \eqref{dot_f0_R} is then:
\begin{equation}\label{neutrino_collission_term_R}
\left(\frac{\partial f}{\partial \tau}\right)_C=3\beta\mathcal{H}\,\frac{\epsilon^2}{\varepsilon^2}\left(1-\frac{q^2}{3\epsilon^2}\right)\left(1-\frac{q^2}{\epsilon^2}\right)\,f\,,
\end{equation}
therefore, the coupling terms of the conservation equations are
\begin{subequations}\label{Q_massive_neutrinos_R}
\begin{align}
Q_{0 h}=&\,4\pi a^{-4}\int q^2\,dq\,\epsilon\,\dot f_0(q,\tau)\,,\\
Q_{1 h}=&\,4\pi a^{-4}\int q^2\,dq\,\epsilon\,\dot f_0(q,\tau)\,\Psi_0\,,\\
Q_{2 h}=&\,4\pi k a^{-4}\int q^2\,dq\,\dot f_0(q,\tau)\,\Psi_1\,.
\end{align}
\end{subequations}

Substituting these last expressions in the evolution equation of vacuum perturbation \eqref{dot_split},  which can be rewritten for the vacuum component associated to massive neutrinos as follow
\begin{subequations}\label{dot_delta_lambda_massive_neutrinos_R}
\begin{align}
\,\overline\rho_{\Lambda h}\left(\dot\delta_{\Lambda h} + 3\mathcal{H} \delta_{\Lambda h}\right) +3\mathcal{H}\frac{Q_{2 h}}{k^2}&=Q_{0 h}\delta_{\Lambda h}-Q_{1 h}\,,\qquad\qquad\text{or}\\
\dot{\delta\rho}_{\Lambda h} +3\mathcal{H}\frac{Q_{2 h}}{k^2}&=-Q_{1 h}\,,
\end{align}
\end{subequations}
where all the  $Q$'s coefficients vanish during the ultra-relativistic regime.
It does not occur for the model presented in the previous section.
Notice that the eqs. \eqref{Q_massive_neutrinos_R} and \eqref{dot_delta_lambda_massive_neutrinos_R} reduce to the expressions \eqref{collision_Q_terms_H2_R} and \eqref{dot_delta_lambda_R}, respectively, in the particular case of $\,\overline{\mathcal{P}}_i/\,\overline\rho_i =\omega_i=\delta\mathcal{P}_i/\delta\rho_i =$const.
In addition, for $\omega_i=1/3$ (in which case it have $\dot{\,\overline\rho}_{\Lambda i}=Q_{0 i}=Q_{1 i}=Q_{2 i}=\dot f_0=0$) the evolution equations for the perturbations of the associated vacuum become $\dot\delta_{\Lambda i}=0$.

We emphasize that the first order Boltzmann equation for massive neutrinos is reduced to the collisionless form, which can be verified by using the expression \eqref{neutrino_collission_term_R} in Equation \eqref{dot_Psi}.
Finally, the relation between $f_0(q,\tau)$ and $\tilde f_0(q)$ is 
\begin{equation}
\frac{\partial\ln f_0}{\partial\ln q}=\frac{\partial\ln\tilde f_0}{\partial\ln q}+q^2\left(\frac{1}{\epsilon^2}-\frac{1}{\epsilon_0^2}\right)+q^2\frac{(1-6\beta)(1-4\beta)}{(1-3\beta)}\left(\frac{1}{\varepsilon_0^2}-\frac{1}{\varepsilon^2}\right)\,.
\end{equation}

\subsection{Coupling terms for energetic components with $\omega=$constant}

For the case of the energetic components with $\,\overline{\mathcal{P}}_i/\,\overline\rho_i =\omega_i=$const. the equation \eqref{conservacion_0_R} becomes
\begin{equation}\label{conservacion_0_omega_R}
\dot{\,\overline\rho}_i+3\mathcal{H}(1+\omega)\,\overline\rho_i=\frac{3\beta(1-3\,\omega_i)}{1-3\beta(1+\omega_i)}\,(1+\omega_i)\,\mathcal{H}\,\,\overline\rho_i\equiv Q_{0i}\,,
\end{equation}
and integrating it we obtain
\begin{equation}\label{densities_R}
\,\overline\rho_i=\,\overline\rho_i^0\,a^{-3(1-4\beta)(1+\omega_i)/(1-3\beta(1+\omega_i))}\,.
\end{equation}
In order to recover Eq. \eqref{conservacion_0_omega_R} we have that the collision term on the Boltzmann equation \eqref{Boltzmann} must be equal to
\begin{equation}\label{collision_term_R}
\left(\frac{\partial\,f}{\partial\tau}  \right)_C=\frac{3\beta(1+\omega_i)}{1-3\beta(1+\omega_i)}(1-3\,\omega_i)\mathcal{H}\,f\,.
\end{equation}
which leads to the coupling terms
\begin{subequations}\label{collision_Q_terms_H2_R}
\begin{align}\label{collision_Q_terms_H2_Ra}
Q_{0i}=&\frac{3\beta(1+\omega_i)}{1-3\beta(1+\omega_i)}(1-3\,\omega_i)\mathcal{H}\,\overline\rho_i\\\label{collision_Q_terms_H2_Rb}
Q_{1i}=&\frac{3\beta(1+\omega_i)}{1-3\beta(1+\omega_i)}(1-3\,\omega_i)\mathcal{H}\,\overline\rho_i\delta_i=Q_{0i}\,\delta_i\\\label{collision_Q_terms_H2_Rc}
Q_{2i}=&\frac{3\beta(1+\omega_i)^2}{1-3\beta(1+\omega_i)}(1-3\,\omega_i)\mathcal{H}\,\overline\rho_i\theta_i=Q_{0i}\,(1+\omega_i)\theta_i\,.
\end{align}
\end{subequations}
Remarkably, for the ultra-relativistic material components ($\omega_i=1/3$), the coupling terms \eqref{collision_Q_terms_H2_R} vanishes, the mean density \eqref{densities_R} becomes $\beta$-independent and evolves as $\bar\rho\propto a^{-4}$ as usual.
In this case there is no a vacuum component associated with ultra-relativistic species.
This is a big difference with the $\Lambda(H^2)$ model.

In addition, as happened in the $\Lambda(H^2)$ model, the conservation equations for the perturbations of standard material components of the universe remain the same as showed in \cite{Bertschinger}.
This can be seen in equations (\ref{dot_splitb_}, \ref{dot_splitc_}), where the right hand side is identically zero because the relationship \eqref{collision_Q_terms_H2_Ra} between the coupling terms.

On the other hand, the right hand side of the conservation equation for vacuum perturbations \eqref{dot_splita_} does not vanish upon substituting the coupling terms \eqref{collision_Q_terms_H2_R}.
In this case, the conservation equation for each $\Lambda i$ component \eqref{dot_splita_} with $\omega_i\neq1/3$ becomes
\begin{equation}\label{dot_delta_lambda_R}
\dot\delta_{\Lambda i}+3\mathcal{H}\delta_{\Lambda i}+\frac{9(1+\omega_i)^2(1-4\beta)}{1-3\beta(1+\omega_i)}\frac{\mathcal{H}^2}{k^2}\theta_i=\frac{3(1+\omega_i)(1-4\beta)}{1-3\beta(1+\omega_i)}\mathcal{H}\left( \delta_{\Lambda i}-\delta_i\right)\,.
\end{equation}

\subsection{Super-horizon initial conditions}\label{super_R}

In order to set the super-horizon ($k\tau \ll 1$) initial conditions for perturbations, we have to solve the standard equations for material species \cite{Bertschinger}, unmodified by the running, plus the additional vacuum equations, sourced by nonrelativistic species.
Because it is expected that $|\beta|\ll1$, we will model the vacuum contributions as test fluids.

During this era only photons and ultra-relativistic neutrinos contribute to the mean density $\,\overline{\rho}_{\text{T}}=\,\overline{\rho}_\gamma + \,\overline{\rho}_\nu$, because their contributions to the vacuum energy density vanishes, see eq. \eqref{densities_R}.
Under these conditions the expansion rate is $\mathcal{H}=\tau^{-1}$.
From Eq. \eqref{dot_delta_lambda_R} we have that the evolution equations of the perturbed vacuum associated to baryons and CDM ($\omega_c=\omega_b=0$) are:
\begin{subequations}\label{primordial_equations}\begin{align}
\dot\delta_{\Lambda b}+3\mathcal{H}\delta_{\Lambda b}+\frac{9(1-4\beta)}{1-3\beta}\frac{\mathcal{H}^2}{k^2}\theta_b &=\frac{3(1-4\beta)}{1-3\beta}\mathcal{H}\left( \delta_{\Lambda b}-\delta_b\right)\,,\\
\dot\delta_{\Lambda c}+3\mathcal{H}\delta_{\Lambda c} &=\frac{3(1-4\beta)}{1-3\beta}\mathcal{H}\left( \delta_{\Lambda c}-\delta_c\right)\,.
\end{align}\end{subequations}
Using the initial conditions found by Ma and Bertschinger  (Eqs. 96 in \cite{Bertschinger}) we find the solutions for Eqs. \eqref{primordial_equations}:
\begin{subequations}\label{ini_R}\begin{align}
\delta_{\Lambda b}&=\frac{2(1-4\beta)}{(2-3\beta)}\,C_1\,\left(k\,\tau\right)^2\,,\\
\delta_{\Lambda c}&=\frac{3(1-4\beta)}{2(2-3\beta)}\,C_1\,\left(k\,\tau\right)^2\,.
\end{align}\end{subequations}
These initial conditions complete the set found by \cite{Bertschinger} for the material and metric perturbations, which sets the relative amplitude (and super-horizon evolution into the radiation dominated era) between the different perturbations of a given length scale $k$.
The relative amplitude, $C_1$, between the different $k$-modes is given by the primordial spectrum parameterized by the scalar spectral index $n_s$.
Finally, the initial condition for the dark energy contribution coming from massive neutrinos follows from the solution for eq. \eqref{dot_delta_lambda_massive_neutrinos_R} with $Q_{1 h}=Q_{2 h}=0$:
\begin{equation}
\delta\rho_{\Lambda h}=c_h\,,
\end{equation}
where the integration constant $c_h$ can be neglected if compared with the growth modes of the other perturbations.

\subsection{Sub-horizon evolution for vacuum perturbations}\label{sub_R}

The perturbations of the vacuum components for the $\Lambda(R)$ model show a similar behavior to the $\Lambda(H^ 2)$ model.
The biggest difference is the lack of a partial vacuum, and vacuum perturbations, coupled to ultra-relativistic material components.

In the non-relativistic massive neutrino case, the evolution equation \eqref{dot_delta_lambda_R} can be approximated as:
\begin{equation}
a\,\delta_\Lambda'=-3\delta_\Lambda+3\,\frac{1-4\beta}{1-3\beta}\,(\delta_\Lambda-\delta)\,.
\end{equation}
Just after the non-relativistic transition the amplitude of $\delta_\Lambda$ is effectively zero, wich leads to the follow equation
\begin{equation}
a\,\delta_\Lambda'=-3\,\frac{1-4\beta}{1-3\beta}\,\delta\,,
\end{equation}
and then, $\delta_\Lambda\approx -3\,\delta/2$, because $\delta\propto a^r$ with $r\sim 2$.

\section{Impact of the Running in the CMB and Matter spectrum}\label{sec_numerical_results}


The evolution equations for the background quantities and linear perturbations were solved using a modified version of the free access code CLASS \cite{class}.
We use the initial conditions shown in the sections \ref{sec_L(H2)_pert_equations} and \ref{sec_L(R)_pert_equations}, which reduce to the adiabatic case when the free parameter of the model vanishes.
In sections \ref{sec_L(H2)_pert_equations} and \ref{sec_L(R)_pert_equations} we showed that the evolution equations for the material perturbations are not modified by the running, in particular, the CDM equations.
In consequence, the usual transformations between synchronous and Newtonian gauges remain unchanged.

The constraint of the free parameter for each model shown at the end of this section was obtained by using the Monte Carlo Markov chain sampler MontePython \cite{MontePython} and the Planck 2015 data \cite{PlanckData}.

\subsection{Evolution of the density contrast}\label{density_contrast_evolution}

\begin{figure}[!h]
\includegraphics[width=8.3cm]{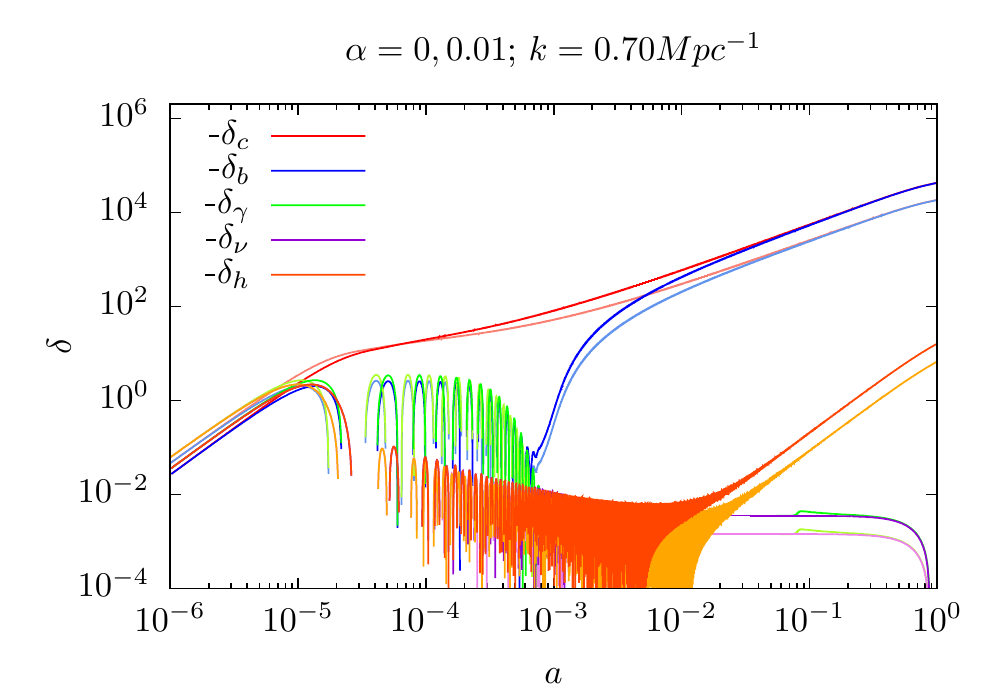}
\includegraphics[width=8.3cm]{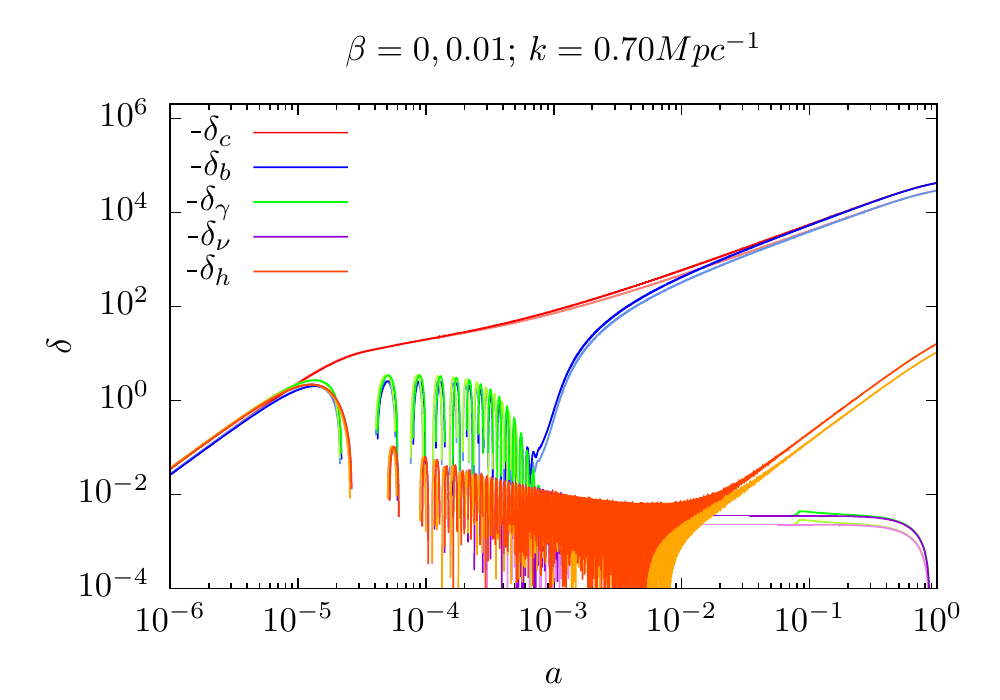}
\caption{Evolution of density contrasts for material components.
{\bf Left panel:} material perturbations with $\Lambda(H^2)$, and $\alpha=\{0,0.01\}$ corresponds to \{bright, light\} colors respectively. {\bf Right panel:} material content with $\Lambda(R)$, where $\beta=\{0,0.01\}$ corresponds to \{bright, light\} colors respectively.
For display purposes $\{\alpha, \beta\}=0.01$ is shown, although, according to observational settings of similar models, it is expected that these parameters are of the order of $10^{-5}-10^{-3}$.}
\label{fig_delta}
\end{figure}

\begin{figure}[!h]
\includegraphics[width=8.3cm]{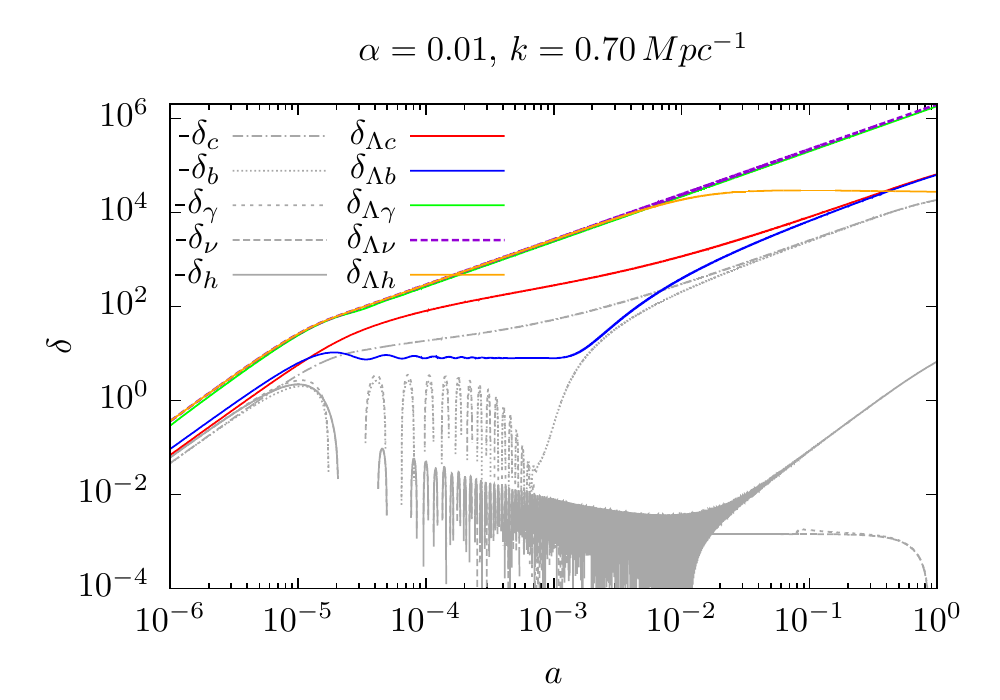}
\includegraphics[width=8.3cm]{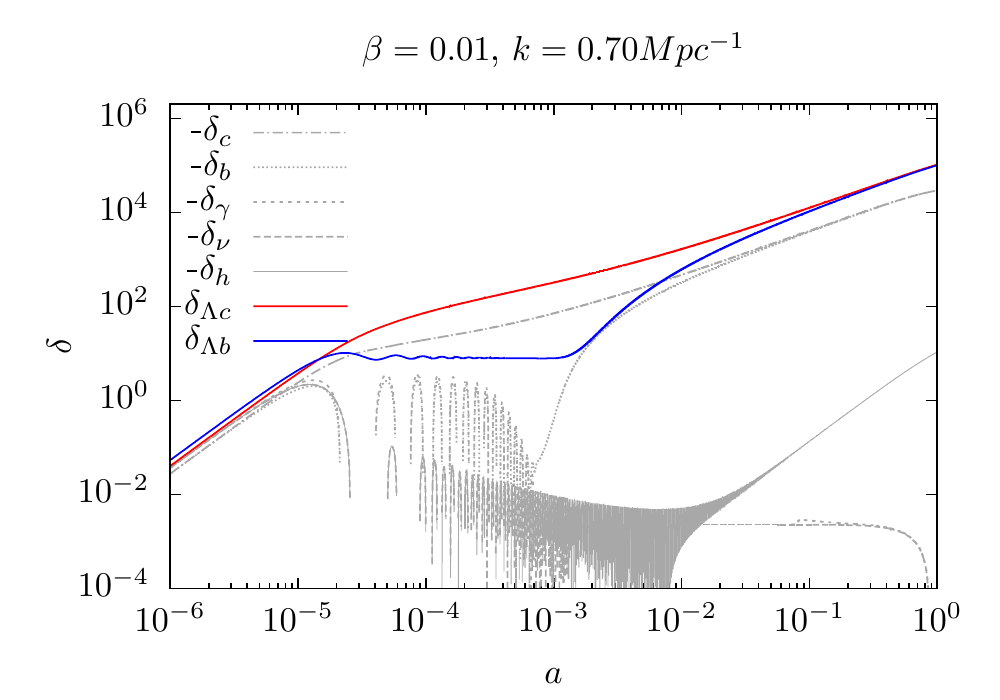}
\caption{Evolution of density contrasts for vacuum components.
{\bf Left panel:} dark energy components with $\Lambda(H^2)$. {\bf Right panel:} dark energy components with $\Lambda(R)$.
For display purposes $\{\alpha, \beta\}=0.01$ is shown, although, according to observational settings of similar models, it is expected that these parameters are of the order of $10^{-5}-10^{-3}$.
Matter perturbations are showed in light gray as a visual reference.}
\label{fig_delta_lambda}
\end{figure}

In Figs. \ref{fig_delta} and \ref{fig_delta_lambda} we show the evolution of the material and vacuum density contrasts, say $\delta_i\equiv\delta \rho_{i}/\,\overline{\rho}_i$ and $\delta_{\Lambda i}\equiv \delta\rho_{\Lambda i}/\,\overline{\rho}_{\Lambda i}$, in both models.
We plot the density contrasts of CDM (red), baryons (blue), photons (green), massless (violet) and massive neutrinos (orange) in Fig. \ref{fig_delta}, and their correspondent partial vacuum contributions in Fig. \ref{fig_delta_lambda}, for the two class of models.

Due to the initial conditions of the $\Lambda(H^2)$ model (light colors), the material density contrasts $\delta_i$ are bigger than the $\Lambda$CDM case (bright colors) outside the horizon, see left panel on Fig. \ref{fig_delta}.
While for superhorizon modes the $\Lambda(R)$ (light colors) model $\delta_i$'s are virtually the same as the  $\Lambda$CDM (bright colors), see right panel on Fig. \ref{fig_delta},  because in this case there is not a coupling between vacuum and ultra-relativistic components, which domain the evolution of both background and perturbation levels.

\subsubsection{Impact of the running {\bf vacuum} on material perturbations}\label{matter_density_contrast_evolution}

The contribution of the vacuum density contrast to the right hand side of the Einstein equation \eqref{Einstein1} is in the form $\delta\rho_{\Lambda i}\propto\alpha\bar\rho_i\delta_{\Lambda i}$ or $\delta\rho_{\Lambda i}\propto\beta\bar\rho_i\delta_{\Lambda i}$.
This vacuum contribution is positive or negative depending on the sign of the running parameter $\alpha$ or $\beta$.
This positive/negative vacuum density contribution increases/reduces the amplitude of the metric perturbation (left-hand side of Einstein equations) in comparison with the standard scenario, where do not have vacuum perturbations.
In consequence, the super-horizon modes of the material density contrast, sourced by metric perturbations, grow slower/faster than those in the $\Lambda$CDM case for positive/negative values of the running parameters $\alpha$ or $\beta$.

Accordingly, this running vacuum effect result in a difference --between the $\Lambda$CDM and the running model-- on the amplitude of the matter and CMB spectra for sub-horizon scales, see the following sections \ref{cmb_spectrum} and \ref{matter_spectrum}.
It is worth noting that the matter and CMB spectrum amplitudes of the running models are larger/smaller than the $\Lambda$CDM case for negative/positive values of $\alpha$ or $\beta$ because the opposite sign between material $\delta_i$ and vacuum $\delta_{\Lambda i}$ density contrasts.

\subsubsection{Vacuum perturbations evolution}

The sub-horizon evolution of the vacuum perturbations was broadly explained as made in sections \ref{sub_H2} and \ref{sub_R}.
Additional features can be seen upon the numerical integration:

\textit{Massive neutrinos}: The mean energy density for massive neutrinos, affected by the running vacuum, decrease slower/faster than the standard $\Lambda$CDM case for positive/negative values of $\alpha$ or $\beta$.
Therefore, the non-relativistic transition is also slower/faster, which delays/anticipate the onset of the non-relativistic regime and beginning of the growth for $\delta_h$ in these cases.

\textit{Baryons}: While the baryon-photon thigh coupling stage, the vacuum perturbations are affected by the baryon oscillations.
After the baryon-photon decoupling, the perturbations of the baryonic vacuum tend to approach the $\delta_{\Lambda c}$, because they evolution equations becomes almost identical.

Remarkably, the bigger amplitude for the vacuum density contrast, relative to material components, does not represent an observational problem for the model, since their impact into the Einstein's equations is suppressed by a $\alpha$ of $\beta$ factor.

\subsection{Matter power spectrum}\label{matter_spectrum}

\begin{figure}[!h]
\includegraphics[width=8.3cm]{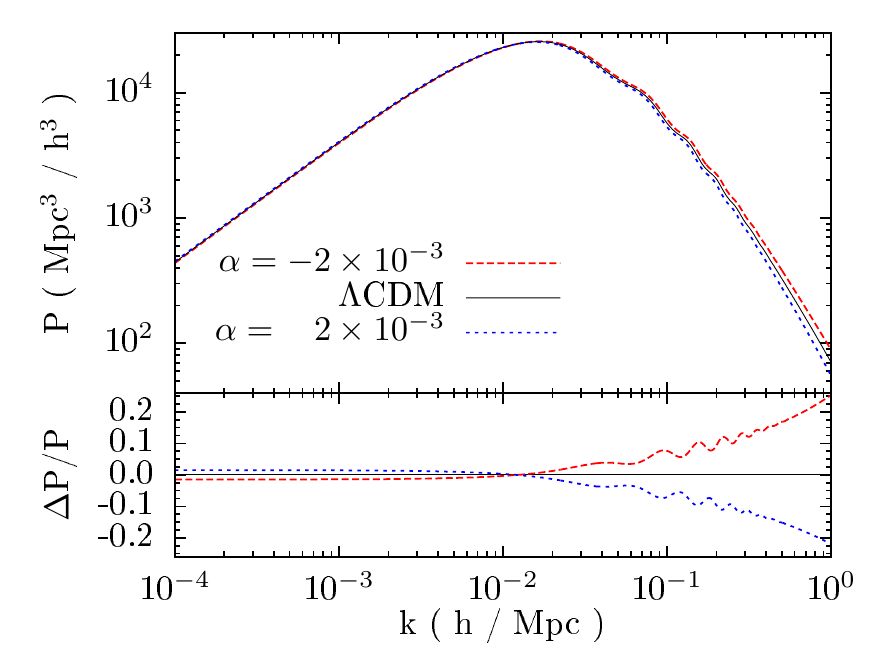}
\includegraphics[width=8.3cm]{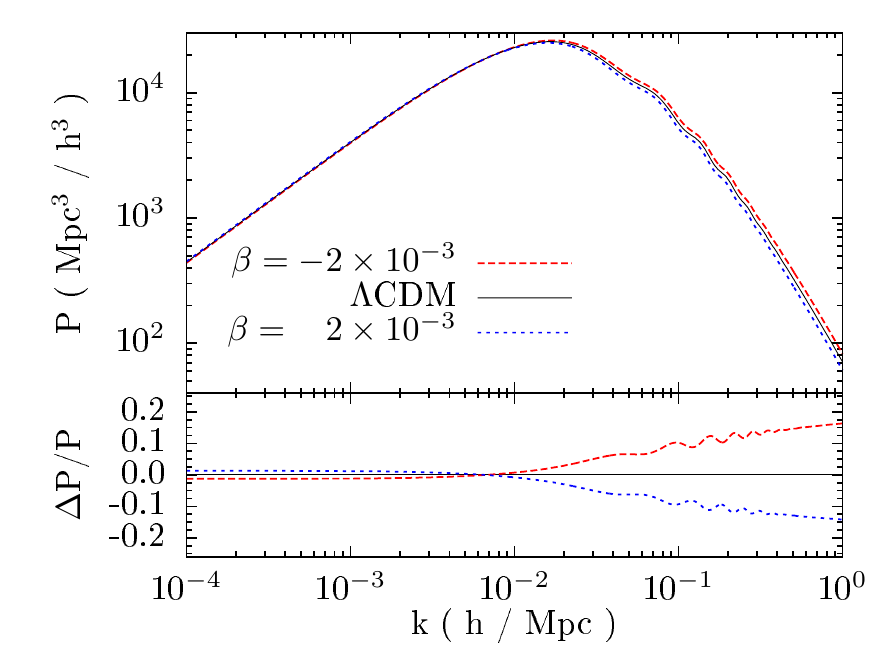}
\caption{Matter power spectrum for the $\Lambda$CDM, $\Lambda(H^2)$ and $\Lambda(R)$ models.
{\bf Left panel:} Comparison between standard and $\Lambda(H^2)$ models.
{\bf Right panel:} Comparison between standard and $\Lambda(R)$ models.
The difference $\Delta P/P$, between the decaying models and the standard model, show distinguishable imprints of the $\Lambda(H^2)$ and the $\Lambda(R)$ models on small scales.
}
\label{fig_pk}
\end{figure}

In Fig. \ref{fig_pk} is shown the matter power spectrum for the $\Lambda(H^2)$ and $\Lambda(R)$ models, in the left and right-hand side respectively.
As discussed in sections \ref{super_H2} and \ref{super_R}, the super-horizon equations for matter perturbations and its initial conditions are the same than in the standard case.
For this reason, the matter power spectrum on large scales is slightly modified by the running.
Even so, the running vacuum induces a lightly and almost scale independent factor due to the difference in the time of transition between the radiation and matter dominated eras $\tau_{eq}$.
When $\alpha\gtrless0$ or $\beta\gtrless0$ we have that $\tau_{eq}|_{\text{running vacuum}}\lessgtr\tau_{eq}|_{\Lambda\text{CDM}}$, therefore the matter perturbations have more time to grow until today, producing an increase/decrease on the matter power spectrum due to the running effects.

On the other hand, as discussed in section \ref{matter_density_contrast_evolution}, the sub-horizon evolution of the matter density contrast --sourced by the metric perturbations--  growth slower/faster for positive/negative values of $\alpha$ or $\beta$.
This results in a lack/excess of the power of matter on small scales, which can be seen in Fig. \ref{fig_pk}.
This difference grows as $k$ increases because the smaller scales come earlier into the horizon, where they are affected for a longer time by the running vacuum.

In brief, in large scale, the matter power spectrum of both running vacuum models are almost indistinguishable.
A possible signature of these models should be in the small scale regime, where the matter power spectrum for $\Lambda(H^2)$ and $\Lambda(R)$ models are distinguishable from each other, see Fig. \ref{fig_pk}.

In addition, the approach presented in this work does not lead to a matter power spectrum instability on small scales for negative values of $\alpha$ or $\beta$, as was found by \cite{Toribio2012,Gomez-Valent2015b,GengLee2016b,GengLee2016b} with the use another formalism for the treatment of linear perturbations.
In the present work, all the observables are affected in a linear way by the small running coefficients $\alpha$ or $\beta$, and reduces to the $\Lambda$CDM case when these parameters vanish.

\subsection{CMB power spectrum}\label{cmb_spectrum}

\begin{figure}[!h]
\hspace{-0.5cm}
\includegraphics[width=8.5cm]{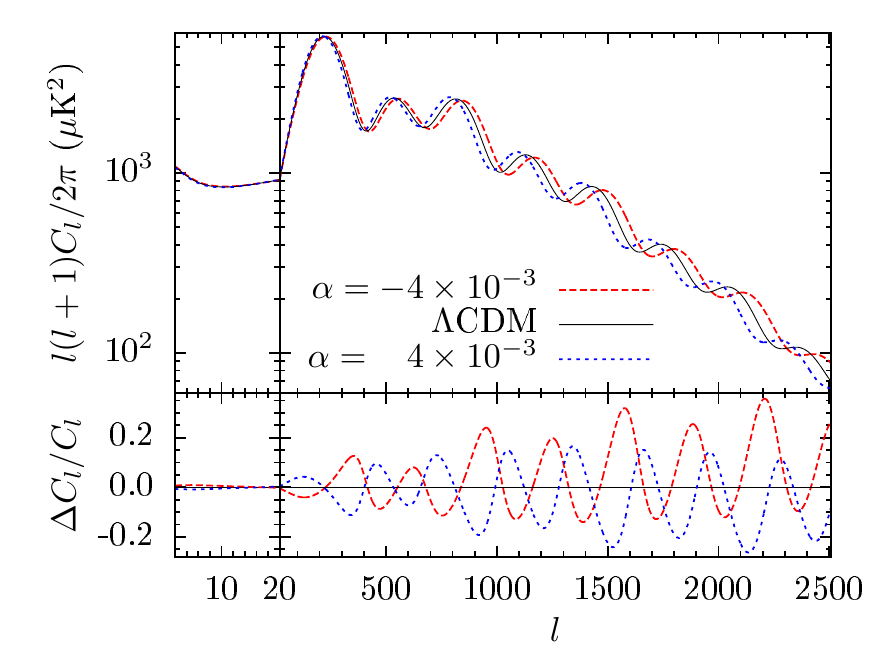}
\hspace{-0.5cm}
\includegraphics[width=8.5cm]{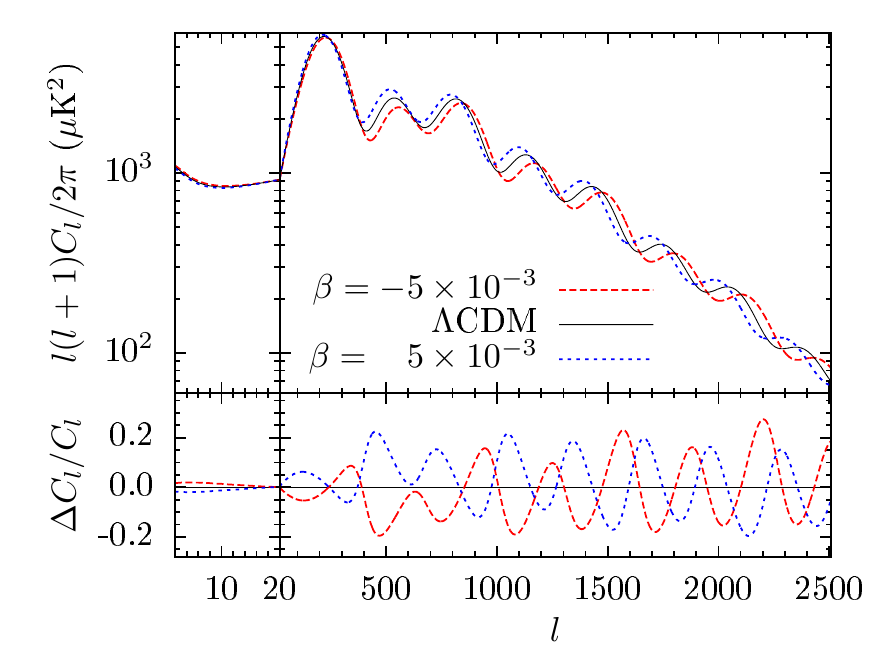}
\caption{CMB temperature anisotropies for the $\Lambda$CDM, $\Lambda(H^2)$ and $\Lambda(R)$ models.
{\bf Left panel:} Comparison between standard and $\Lambda(H^2)$ models.
{\bf Right panel:} Comparison between standard and $\Lambda(R)$ models.
}
\label{fig_cl}
\end{figure}

The CMB temperature anisotropies of the $\Lambda$CDM, $\Lambda(H^2)$ and $\Lambda(R)$ models are presented.
Previous parameter constraints suggest $\alpha, \beta \sim 10^{-5}-10^{-3}$ \cite{SolaAIP14, Gomez-Valent:2015pia, Sola:2015wwa}.
In Fig. \ref{fig_cl} we show the $TT$ spectrum of the CMB anisotropies, for $\Lambda(H^2)$ in the left side, and $\Lambda(R)$ in the right side.
Below we describe the most relevant effects of the running vacuum on the $TT$ power spectrum of the CMB, which implies some degeneracies between the basic cosmological parameters and the running parameters, $\alpha$ or $\beta$.
\subsubsection{Peak positions}
In the case of the $\Lambda(H^2)$ model, left panel of Fig. \ref{fig_cl}, the main effect of the running seems to be that all peak positions of the $TT$ spectrum are shifted to low/high $l$'s when the free parameter $\alpha$ assumes positive/negative values.
The peak shift seem proportional to the reference $\Lambda$CDM position $l$ and to the value of the parameter $\alpha$.
That effect is too present in the $\Lambda(R)$ running model, right panel of Fig. \ref{fig_cl}, but the shift is less sensitive to the variation of $\beta$.
In the context of the $\Lambda$CDM model this effect can be compensated principally by the physical densities of cold dark matter and baryons today, $\Omega_ch^2$ and $\Omega_bh^2$ respectively.
Besides, the bottom part of the Fig \ref{fig_cl} show that the difference in the amplitude of the spectrum is not the same for odd and even peaks for both running models.
This effect is in the same way that the produced by the variation of the $\Omega_bh^ 2$, which implies a degeneracy between this parameter and the parameters $\alpha$ and $\beta$ of the running models

On the other hand, the change on the angular size of the sound horizon at recombination $\theta_*$ also shift the peak positions, but at the same time change the amplitude of the power spectrum for $l\lesssim10$.
Since that last effect is not a consequence of the running vacuum, see Fig. \ref{fig_cl}, it is not expected a degeneracy between $\theta_*$ and the running parameters $\alpha$ or $\beta$.

\subsubsection{Amplitude}
The stronger effect of the running for the model $\Lambda(R)$ is the increase/decrease of the power amplitude for positive/negative values of $\beta$, see right panel of Fig. \ref{fig_cl}.
This change occurs in the same way for almost all the scales of the $TT$ spectrum, except for the low multipoles, where the running effect is weaker.
The $\Lambda(H^2)$ model present the same kind of variation with $\alpha$, but in this case, the effect is smaller.
In the case of the $\Lambda(R)$ model, the variation of the optical depth during reionization  $\tau_{reio}$, should help to compensate the running effect; as this cosmological parameter modify the amplitude of the $TT$ spectrum after the fist peak.

In addition, in the Fig. \ref{fig_cl}, it can see that a global compensation on the amplitude is not enough to correct the running effects of the model $\Lambda(H^2)$ since their discrepancy with the standard model grows with the index $l$ of the multipoles.
This monotonous discrepancy in the amplitude can be countered by the variation of the scalar spectral index value $n_s$.

\subsubsection{Main degeneracies}
These running vacuum effects, in the temperature spectrum of the CMB, would be the main sources of correlation between $\alpha$ and the cosmological parameters $\{\Omega_bh^ 2\,,\,\tau_{reio}\,,\,n_s\}$ in the case of the $\Lambda(H^2)$ running model, and between $\beta$ and $\{\tau_{reio}\,,\,\Omega_bh^ 2\}$ for the $\Lambda(R)$ model.

The $\beta$ degeneracy with the baryon fraction $\Omega_b$ arises because the baryon-photon ratio, for the $\Lambda(R)$ running model, is different that it in the standard model for all redshift different to zero:
\begin{equation}
\left.\frac{\bar\rho_b}{\bar\rho_\gamma}\right|_{\Lambda(R)}=\frac{\Omega_b}{\Omega_\gamma}\,a^{1/(1-3\beta)}\lessgtr\frac{\Omega_b}{\Omega_\gamma}\,a=\left.\frac{\bar\rho_b}{\bar\rho_\gamma}\right|_{\Lambda\text{CDM}}\,,\quad \text{for }\beta\gtrless0\qquad\text{and }a<1\,.
\end{equation}
The $\Lambda(H^2)$ is affected for the running vacuum in the same way, but with a weaker and opposite dependence on the parameter $\alpha$:
\begin{equation}
\left.\frac{\bar\rho_b}{\bar\rho_\gamma}\right|_{\Lambda(H^2)}=\frac{\Omega_b}{\Omega_\gamma}\,a^{1-\alpha}\gtrless\frac{\Omega_b}{\Omega_\gamma}\,a=\left.\frac{\bar\rho_b}{\bar\rho_\gamma}\right|_{\Lambda\text{CDM}}\,,\quad \text{for }\alpha\gtrless0\qquad\text{and }a<1\,.
\end{equation}

The $\{\tau_{reio},\,n_s \}$ degeneracy with the running vacuum  parameters $\alpha$ or $\beta$ --related to the spectrum amplitude-- has explanation out of the background approach: the material perturbations growth faster/slower in the context of the running vacuum model than within the standard cosmological scenario, for a negative/positive values of $\alpha$ or $\beta$, see section \ref{density_contrast_evolution}.

Finally, it is worth nothing that for low multipoles, $l \leq 20$, the temperature spectrum are very little affected by the running of vacuum.
This occurs because the super-horizon evolution of the material perturbations is the same in the standard and running models.
Therefore, the larger physical scales are poorly affected by the running.

\subsection{Planck constraints}

\begin{figure}[!h]
\includegraphics[width=17cm]{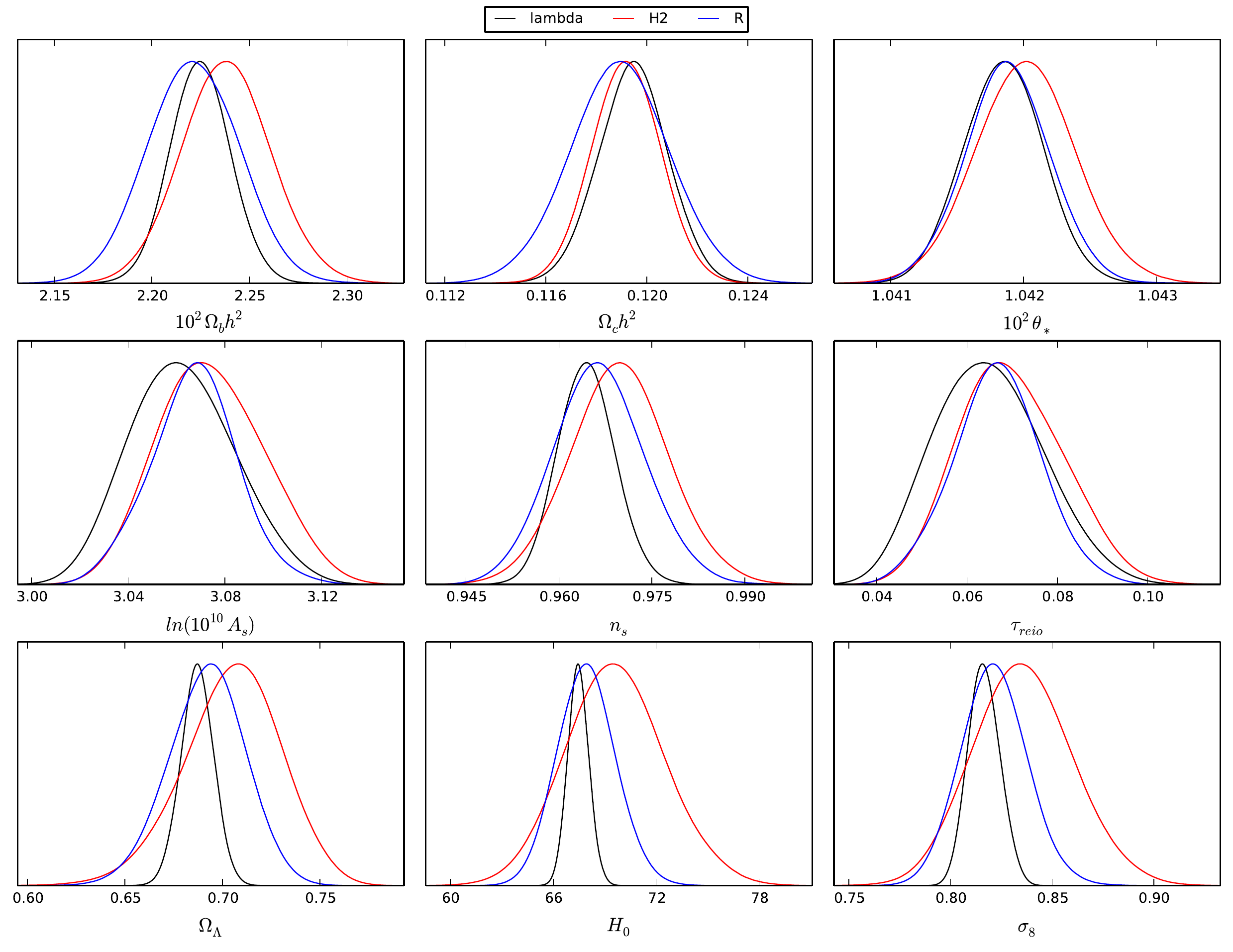}
\caption{Comparison between the posterior distribution of the cosmological parameters of the $\Lambda(H^2)$ and $\Lambda(R)$ running vacuum models and the six-parameter cosmological standard model and the running vacuum model $\Lambda(R)$.
$\{\Omega_bh^2\,,\,\Omega_ch^2\,,\,\theta_*\,,\,A_s\,,\,n_s\,,\,\tau_{reio} \}$ are set of input cosmological parameters, while $\{\Omega_\Lambda\,,\,H_0\,,\,\sigma_8 \}$ are calculated parameters.
$H_0$ units are $km\,s^{-1}\,Mpc^{-1}$.}
\label{fig_H2-R-lambda}
\end{figure}

\begin{figure}[!h]
\includegraphics[width=17cm]{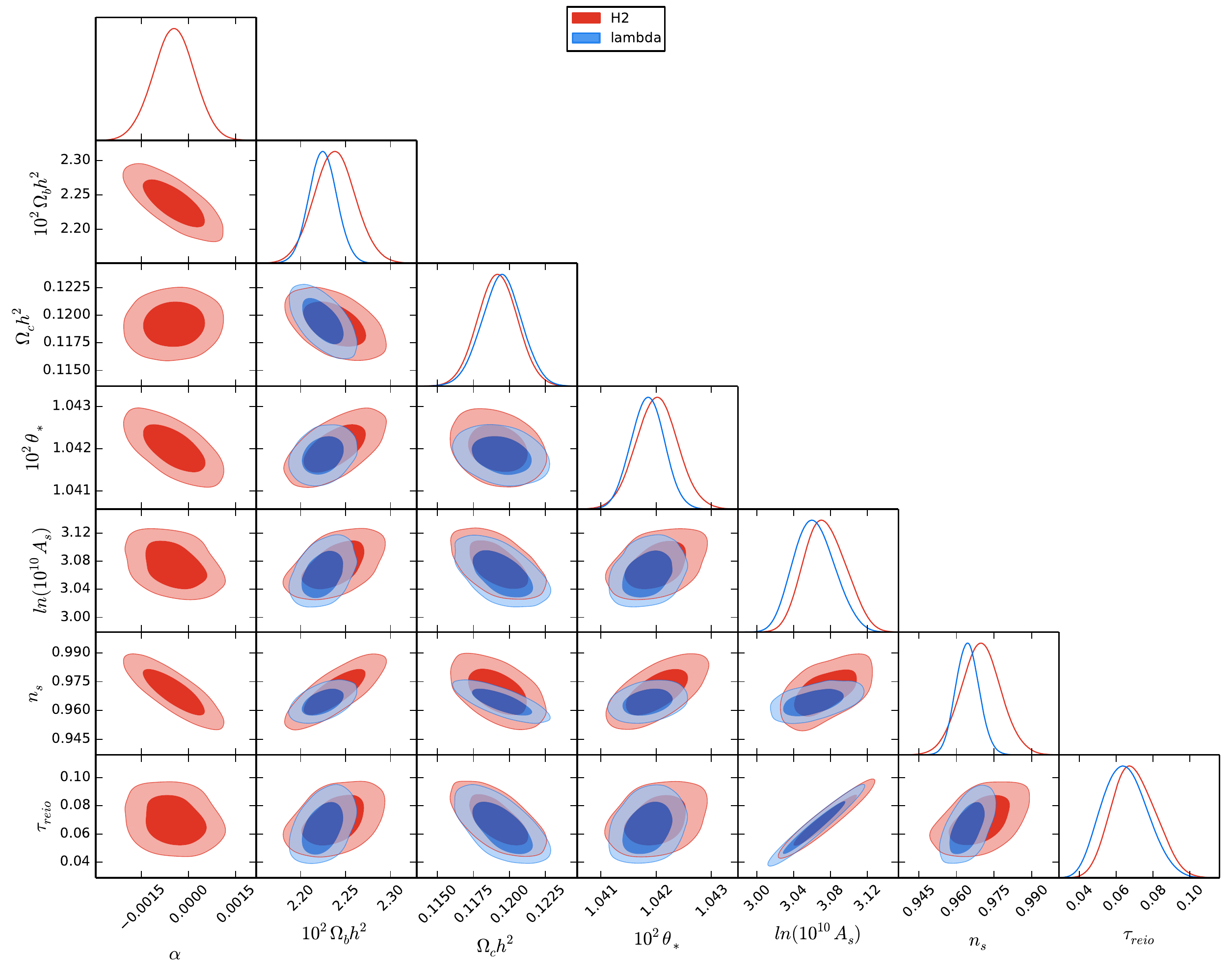}
\caption{Comparison between the CBM anisotropies constrain of the six-parameter cosmological standard model and the running vacuum model $\Lambda(H^2)$.}
\label{fig_H2-lambda}
\end{figure}

In order to constrain the free parameters of the $\Lambda(H^2)$ and $\Lambda(R)$ models, we confront the theory with the observational data given by the CMB anisotropy spectrum.
The theory prediction was calculated by integrating numerically the complete set of evolution equations for both the background and the scalar linear perturbations using the code CLASS \cite{class}, which was modified in order to include the vacuum perturbations and the background effects of the running.
The observational data used here was the high and low $C_l$'s corresponding to the $TT$, $TE$ and $EE$ CMB spectrum given by the 2015 data release of the Plank Collaboration \cite{PlanckData}.
Monte Carlo Markov Chains were generated using the MontePython free code \cite{MontePython}, while the statistical analysis and plots were carried out by using the GetDist package \cite{GetDist}.
Additionally, in order to establish a fair comparison between the six-parameter cosmological standard model and the $\Lambda(H^2)$ or $\Lambda(R)$ models, it was fixed the effective numbers of neutrinos $N_{eff}=3.046$, the neutrino mass $m_\nu=0.06$eV and a single massive neutrino species, as was done by the Planck Collaboration \cite{PlanckCosmology}.
The results are summarized in Table \ref{tabla}, and Figs. \ref{fig_H2-R-lambda}, \ref{fig_H2-lambda} and \ref{fig_R-lambda}.
\begin{table}[!h]
\begin{tabular}{llllllllll}
\hline 
 & \multicolumn{3}{c}{$\Lambda$CDM} & \multicolumn{3}{c}{$\Lambda(H^2)$}& \multicolumn{3}{c}{$\Lambda(R)$}\\ \cline{2-10}
Param$\quad$ & best-fit$\;\;$ & mean &$\pm\sigma\qquad\quad$& best-fit$\;\;$ & mean&$\pm\sigma\qquad\quad$ & best-fit$\;\;$ & mean&$\pm\sigma$ \\ \hline  \hline 
$10^{4}\alpha$ &\;\;---&\;\;---&\;\;---&$1.71$ & -$4.7$&$\pm 6.5$ &\;\;---&\;\;---&\;\;--- \\ 
$10^{4}\beta$ &\;\;---&\;\;---&\;\;---&\;\;--- &\;\;---&\;\;---&-$1.44$ & -$1.4$&$\pm 5.6$ \\  \hline 
$10^2\Omega_{b }h^2$ &$2.223$ & $2.225$&$\pm 0.015$ &$2.208$ & $2.238$&$\pm 0.023$ & $2.222$ & $2.221$&$\pm 0.024$\\ 
$10~\Omega_{c }h^2$ &$1.120$ & $1.195$&$\pm 0.014$ & $1.206$ & $1.192$&$\pm 0.014$ & $1.187$ & $1.189$&$\pm 0.020$\\ 
$100~\theta_{* }$ &$1.0418$ & $1.04185$&$\pm 0.00030$ & $1.0418$ & $1.04201$&$\pm 0.00038$ & $1.04193$ & $1.04188$&$\pm 0.00032$\\
$ln(10^{10}A_{s } )$ &$3.044$ & $3.062$&$^{+0.021}_{-0.024}$ & $3.050$ & $3.074$&$^{+0.021}_{-0.024}$ & $3.078$ & $3.067$&$\pm 0.018$\\  
$n_{s }$ &$9.619$ & $9.644$&$\pm 0.046$ & $9.562$ & $9.697$&$\pm 0.080$ & $9.688$ & $9.664$&$\pm 0.074$\\ 
$10^2\tau_{reio }$ &$5.351$ & $6.5$&$^{+1.1}_{-1.3}$ & $5.802$ & $7.0$&$^{+1.1}_{-1.3}$ & $7.257$ & $6.67$&$\pm 0.97$\\  \hline
$10~\Omega_{\Lambda }$ &$6.842$ & $6.871$&$\pm 0.085$ & $6.732$ & $7.05$&$^{+0.27}_{-0.23}$ & $6.943$ & $6.92$&$\pm 0.20$\\ 
$H_{0 }$ &$67.26$ & $67.47$&$\pm 0.62$ & $66.24$ & $69.6$&$\pm 2.9$ & $68.05$ & $67.9$&$\pm 1.7$\\ 
$10~\sigma_{8 }$ &$8.099$ & $8.165$&$\pm 0.078$ & $8.087$ & $8.36$&$\pm 0.25$ & $8.258$& $8.21$&$\pm 0.16$\\  
\hline 
\end{tabular}
\caption{Best fit and confidence level list for the cosmological parameters of the $\Lambda$CDM, $\Lambda(H^2)$ and $\Lambda(R)$ models given by the $TT,TE,EE+lowP+lensing$ statistical analysis of the Planck 2015 data \cite{PlanckData}.} \label{tabla}
\end{table}

\subsubsection{Main parameters}

\begin{figure}[!h]
\includegraphics[width=17cm]{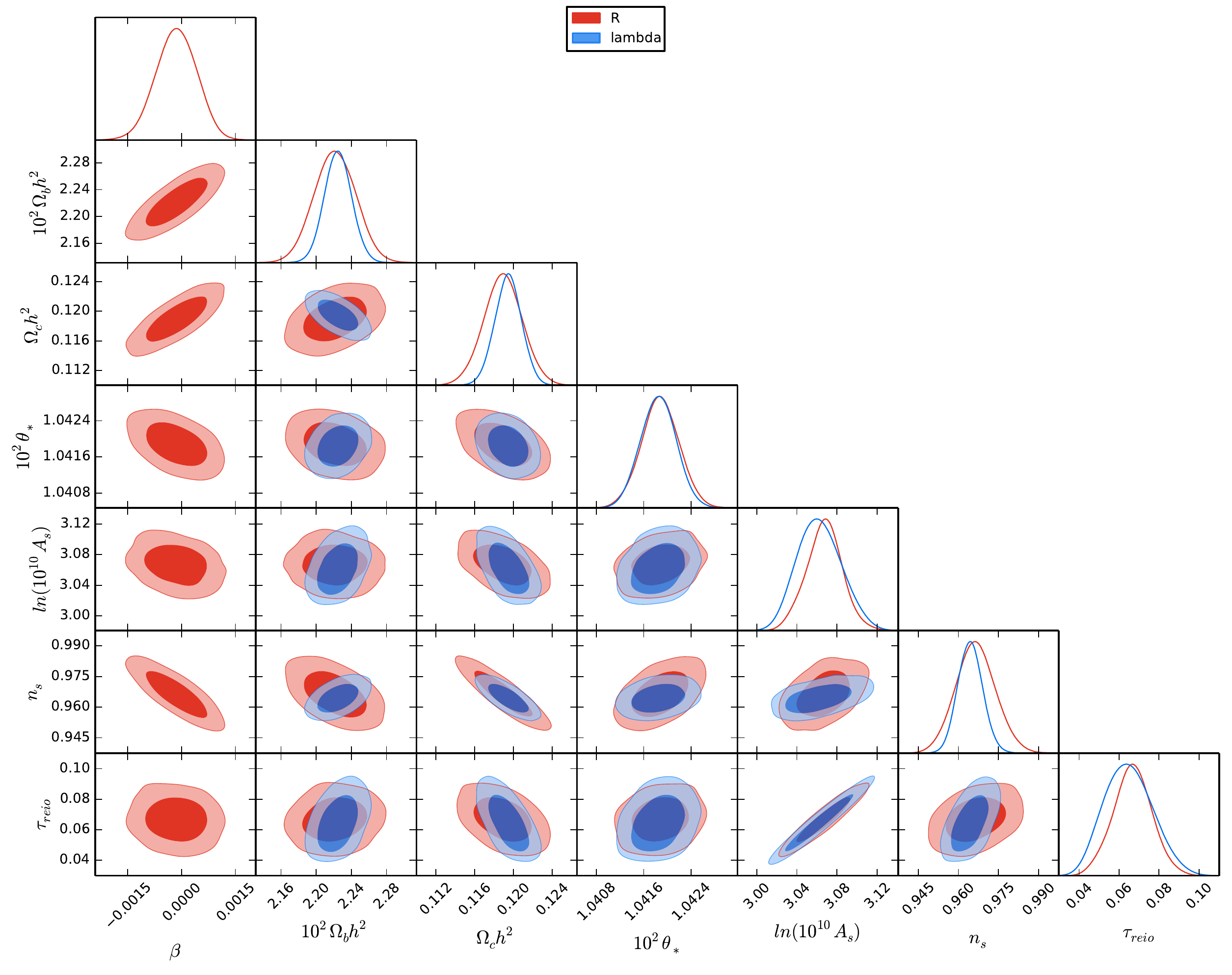}
\caption{Comparison between the CBM anisotropies constrain of the six-parameter cosmological standard model and the running vacuum model $\Lambda(R)$.}
\label{fig_R-lambda}
\end{figure}

The data constraints of the CMB power spectrum, Table \ref{tabla} and Figs. \ref{fig_H2-R-lambda} and \ref{fig_R-lambda}, show that the cosmological parameters more affected by the $\Lambda(H^2)$ model are $\{n_s\,,\,\Omega_bh^2\,,\,\theta_*\,,\,ln(10^{10}A_{s })\,,\,\tau_{reio}\}$, whose means was shifted $\{1.2\sigma\,,\,1\sigma\,,\,0.5\sigma\,,\,0.5\sigma\,,\,0.4\sigma\}$ respectively regarding the $\Lambda$CDM constraints.

A larger value of the spectral index $n_s$ can offset the monotonic decrease of amplitude, as a function on $l$, generated by the running vacuum on high multipoles, with respect to the $\alpha=0$ case, see Fig. \ref{fig_cl}.
A larger value of the baryon fraction $\Omega_bh^2$ generates roughly a systematic difference in amplitude between even and odd peaks of the spectrum, in the same way as that produced by the running, see bottom box on Fig. \ref{fig_cl}.
At the same time, the larger value of $\Omega_bh^2$ shift the peak positions to the right, increasing the discrepancy generated by the running vacuum effects.
This shift in the peak positions can be offset with a larger value of $\theta_*$.
The residue of the $n_s$ correction on the spectrum amplitude is then in hands of $A_s$ and $\tau_{reio}$.
$A_s$ is responsible by the global amplitude of the spectrum, while roughly $\tau_{reio}$ only the multipoles greater than $l\sim100$.  

In the case of the $\Lambda(R)$ model, the parameters $\{\Omega_ch^2\,,\,n_s\,,\,\Omega_bh^2\,,\,ln(10^{10}A_{s })\,,\,\tau_{reio}\,\}$ shifted $\{-0.43\sigma\,,\,0.43\sigma\,,\,-0.27\sigma\,,\,0.21\sigma\,,\,0.13\sigma\, \}$ respecting to the $\Lambda$CDM constraints.
In comparison with the $\Lambda(H^2)$ model, the running effect of the $\Lambda(R)$ model produces a bigger reduction of the height of the first peaks (for negative values of $\beta$, see Fig. \ref{fig_cl}).
This effect can be compensated by smaller values of the cold dark matter fraction $\Omega_ch^2$.
A lower value of $\Omega_ch^ 2$ also shift the peaks position of the spectrum toward lower values of $l$, dispensing the change on $\theta_s$, which leaves almost unchanged the height of the peaks.

\subsubsection{Calculated parameters}
According with table \ref{tabla}, see also Fig. \ref{fig_H2-R-lambda}, the calculated cosmological parameters $\{\,\Omega_\Lambda\,,\,H_0\,,\,\sigma_8 \}$ are shifted to higher values in the presence of the running vacuum $\Lambda(R)$, with respect to the standard model.
In the case of the $\Lambda(R)$ model, the mean of this cosmological parameters $\{\,\Omega_\Lambda\,,\,H_0\,,\,\sigma_8 \}$ increase in $\{\,0.6\,\sigma\,,\,0.9\,\sigma\,,\,0.6\,\sigma\}$, but their variances grow by a factor of $\{\,2.4\,,\,2.7\,,\,2.3\}$, with respect to the $\Lambda$CDM values.

For the $\Lambda(H^2)$ model, the mean of the calculated cosmological parameters $\{\,\Omega_\Lambda\,,\,H_0\,,\,\sigma_8 \}$ increase in $\{\,2.1\,\sigma\,,\,3.4\,\sigma\,,\,2.5\,\sigma\}$ while their variances grow by a factor of $\{\,3.2\,,\,4.7\,,\,3.2\}$, with respect to the $\Lambda$CDM values.

The positive shift in the calculated cosmological parameters $\{\,\Omega_\Lambda\,,\,H_0\,,\,\sigma_8 \}$, due to the running vacuum effects, is in the same direction of low-redshift constraints of the $\Lambda$CDM cosmological model.
In this sense, the running vacuum models could reduce the tension between the hight redshift and the low observational constraints of $\Omega_\Lambda$ and $H_0$.
In addition, because the background densities and geometric cosmological quantities of both the running vacuum models have the same values as the standard model at redshift zero, tough their evolution are affected by the running, it is expected that low redshift cosmological tests of the running models may to improve the constraints of $\Omega_\Lambda$, $H_0$ and $\alpha$ or $\beta$ parameters.

\section{Conclusions}\label{sec_conclusions}

This paper seeks to advance the linear perturbations study of phenomenological models for a running cosmological constant built from the main ideas of the renormalization group.
We generalized two running vacuum modeling, considering the energy exchange between the vacuum and each material species, not only with the dominant component as is often done in interacting dark sector models.
This improvement in such approach is necessary in order to analyze the effects of the model on perturbations.

We considered two representative classes models, $\Lambda(H^2)$ and $\Lambda(R)$, derived and solved numerically the perturbation equations.
In addition, we plotted two important cosmological observables: CMB temperature and matter power spectrum; as well as the evolution of the perturbations.
The hypothesis of vacuum perturbations is viable if we consider a dynamical vacuum modeled as a perfect fluid.
This could possibly indicate which observables are the most appropriate to track with the purpose to give more conclusive observational supports to the models. 
Although the both $\Lambda(H^2)$ and $\Lambda(R)$ models studied in this paper lacking of a Lagrangian formulation, the use of the first order collision terms inherited from the Boltzmann ones to zero order allows a self-consistent treatment for perturbations.
As a result, the matter perturbations are only affected by its vacuum counterparts through the metric perturbations, leaving unmodified the usual transformation between synchronous and Newtonian gauges.
Unlike the material components, the equations for vacuum perturbations have a non-zero collision term.

Besides being able to consistently formulate a set of equations for the evolution of vacuum perturbations, in the context of the model presented here, we have put them in the middle of the cosmic inventory and estimate their impact on material components.
In principle, the contribution of vacuum perturbations to the gravitational potential, throughout the cosmic history, allow that the small scales coming on the horizon are the most affected by the interaction.
The numerical solutions show that this effect results in a decrease/increase of matter power spectrum in small scales, also as a leftward/rightward shift on the peak positions in the $TT$ spectrum of the CMB depending on the sign of the running parameters $\alpha$ and $\beta$.
In addition, the vacuum coupling also modifies the slope of the matter spectrum on large scales; which becomes a potential source of observational distinction between the two models.

The study developed in this work, complements similar ones concerning running vacuum, vacuum scalar perturbations and dark sector interaction; establishing a detailed derivation of the linear scalar perturbation in two well-motivated classes of running vacuum ($\Lambda(H^2)$ and $\Lambda(R)$), which interacts with all the material species.
The vacuum perturbations were tracked in detail.

The Planck constraints obtained in this work for the $\Lambda(H^2)$ model gives $10^4\alpha=-4.7\pm 6.5$, which is consistent with the $\Lambda$CDM model in $\sim0.72\sigma$, while $10^4\beta=-1.4\pm5.6$ is almost indistinguishable of the the $\Lambda$CDM model.
Remarkable, it has that $\Lambda(H^2)$ calculated parameters constrained by the CMB power spectrum, $\Omega_\Lambda$ and $H_0$, seems to alleviate the tension with low-redshift observations: they present a positive sifts with respect to the standard case.
Even so, the sifts in this parameters --due to the vacuum running-- are compatible with the $\Lambda$CDM case, because the constrains becomes weaker due to the degeneracies between the running parameters and the standard cosmological parameters.
The enhancement of the statistical analysis with the use of additional observations such as BAO, SNe and $H(z)$ measurements could both improve the constraints in $\Omega_\Lambda$ and $H_0$, and confirm the reduction of the low and hight redshift stress present in the $\Lambda$CDM studies.

\subsubsection*{Acknowledgements}
ELDP. is supported by FAPESP under grants 2015/01721-9.
DAT was partially supported by CAPES.
This work has made use of the computing facilities of the Laboratory of Astroinformatics (IAG/USP, NAT/Unicsul), whose purchase was made possible by the Brazilian agency FAPESP (grant 2009/54006-4) and the INCT-A.

\bibliography{refs_art_Perico_Tamayo}
\bibliographystyle{unsrt}

\end{document}